\documentclass{article}

\usepackage{microtype}
\usepackage{graphicx}
\usepackage{booktabs} %
\usepackage{xcolor}
\usepackage{titletoc}
\usepackage{amsmath}
\usepackage{amsthm}
\usepackage{amssymb}
\usepackage{hyperref}
\usepackage{siunitx}
\usepackage{lipsum}
\usepackage[inline]{enumitem}

\usepackage{multirow}
\usepackage{array}
\usepackage{multicol}
\usepackage{tabularx}

\usepackage[utf8]{inputenc}
\usepackage{booktabs} %
\usepackage{bbm} 
\usepackage{bm}
\usepackage{verbatim}
\usepackage{float}
\usepackage{enumitem}
\usepackage{mathtools}
\usepackage{hhline}
\usepackage[title]{appendix}
\usepackage[nameinlink]{cleveref} %
\usepackage[font=small,labelfont=bf,
   justification=justified,
   format=plain]{caption}
\usepackage{subcaption}
\usepackage{wrapfig,booktabs}
\usepackage{xspace}
\usepackage{cancel}
\usepackage{sidecap}
\DeclarePairedDelimiterX{\infdivx}[2]{(}{)}{%
  #1\;\delimsize\|\;#2%
}

\newcommand{\bnns}{\textsc{bnn}s\xspace}

\newcommand{\real}{\mathbb{R}}

\crefname{appsec}{appendix}{appendices}
\Crefname{appsec}{Appendix}{Appendices}

\newcommand{\bX}{\mathbf X}
\newcommand{\by}{\mathbf y}
\newcommand{\bY}{\mathbf Y}
\newcommand{\bx}{\mathbf x}

\newcommand{\dee}{\,\textrm{d}}

\newcommand{\R}{\mathbb{R}}

\newcommand{\btheta}{{\boldsymbol{\theta}}}

\newcommand{\closer}[3]{{\kern-#1ex{#2}\kern-#3ex}}

\newcommand{\DKL}{\DD_{\textrm{KL}}\infdivx}

\mathchardef\mhyphen="2D

\DeclareMathOperator{\E}{\mathbb{E}}

\usepackage{algorithm}
\usepackage{algorithmic}
\usepackage{colortbl}

\usepackage{siunitx}
\usepackage{standalone}
\newcommand\defines{\,\dot{=}\,}

\newcommand{\vbar}{\,|\,}

\newcommand{\calX}{\mathcal{X}}
\newcommand{\calY}{\mathcal{Y}}
\newcommand{\calD}{\mathcal{D}}

\newcommand{\calQ}{\mathcal{Q}}
\newcommand{\calC}{\mathcal{C}}

\newcommand{\DD}{\mathbb{D}}

\newcommand{\bTheta}{\boldsymbol{\Theta}}

\newcommand{\pms}[1]{\ensuremath{{\scriptstyle\pm #1}}}

\usepackage{tcolorbox}

\newcommand{\codebox}[2]{%
\begin{tcolorbox}[colback=blue!10!white,leftrule=2.5mm,size=title]
#1: #2
\end{tcolorbox}
\vspace{-0.1cm}%
}

\usepackage[algo2e]{algorithm2e}

\newcolumntype{x}[1]{>{\centering\let\newline\\\arraybackslash\hspace{0pt}}p{#1}}

\usepackage[accepted]{include/icml/icml2023}

\icmltitlerunning{
Drug Discovery under Covariate Shift with Domain-Informed Prior Distributions over Functions
}

\begin{document}

\twocolumn[
\icmltitle{
     Drug Discovery under Covariate Shift with\\Domain-Informed Prior Distributions over Functions
}

\icmlsetsymbol{equal}{*}

\begin{icmlauthorlist}
\icmlauthor{
    Leo Klarner
}{oxford}
\icmlauthor{
    Tim G. J. Rudner
}{oxford}
\icmlauthor{
    Michael Reutlinger
}{roche}
\icmlauthor{
    Torsten Schindler
}{roche}
\\\icmlauthor{
    Garrett M. Morris
}{oxford}
\icmlauthor{
    Charlotte M. Deane
}{oxford}
\icmlauthor{
    Yee Whye Teh
}{oxford}
\end{icmlauthorlist}

\icmlaffiliation{oxford}{Department of Statistics, University of Oxford, United Kingdom}
\icmlaffiliation{roche}{Pharma Research \& Early Development, Roche, Switzerland}

\icmlcorrespondingauthor{Leo Klarner}{leo.klarner@stats.ox.ac.uk}

\vskip 0.3in
]

\printAffiliationsAndNotice{}  %
\begin{abstract}
Accelerating the discovery of novel and more effective therapeutics is an important pharmaceutical problem in which deep learning is playing an increasingly significant role. However, real-world drug discovery tasks are often characterized by a scarcity of labeled data and significant covariate shift---a setting that poses a challenge to standard deep learning methods. In this paper, we present \textsc{q-savi}, a probabilistic model able to address these challenges by encoding explicit prior knowledge of the data-generating process into a prior distribution over functions, presenting researchers with a transparent and probabilistically principled way to encode data-driven modeling preferences. Building on a novel, gold-standard bioactivity dataset that facilitates a meaningful comparison of models in an extrapolative regime, we explore different approaches to induce data shift and construct a challenging evaluation setup. We then demonstrate that using \textsc{q-savi} to integrate contextualized prior knowledge of drug-like chemical space into the modeling process affords substantial gains in predictive accuracy and calibration, outperforming a broad range of state-of-the-art self-supervised pre-training and domain adaptation techniques.
\end{abstract}

\section{Introduction}

Discovering novel drug candidates that are able to safely and effectively treat neglected diseases or combat multidrug-resistant pathogens is a challenging biomedical research problem of considerable scientific and societal importance. 
Leveraging modern deep learning algorithms to accurately predict clinically relevant molecular properties and reduce the need for time- and resource-intensive experiments has the potential to significantly accelerate the development of promising and innovative chemical leads in drug discovery.
\begin{figure}[!t]
\centering
    \vspace{20pt}
    \includegraphics[width=\columnwidth, ]{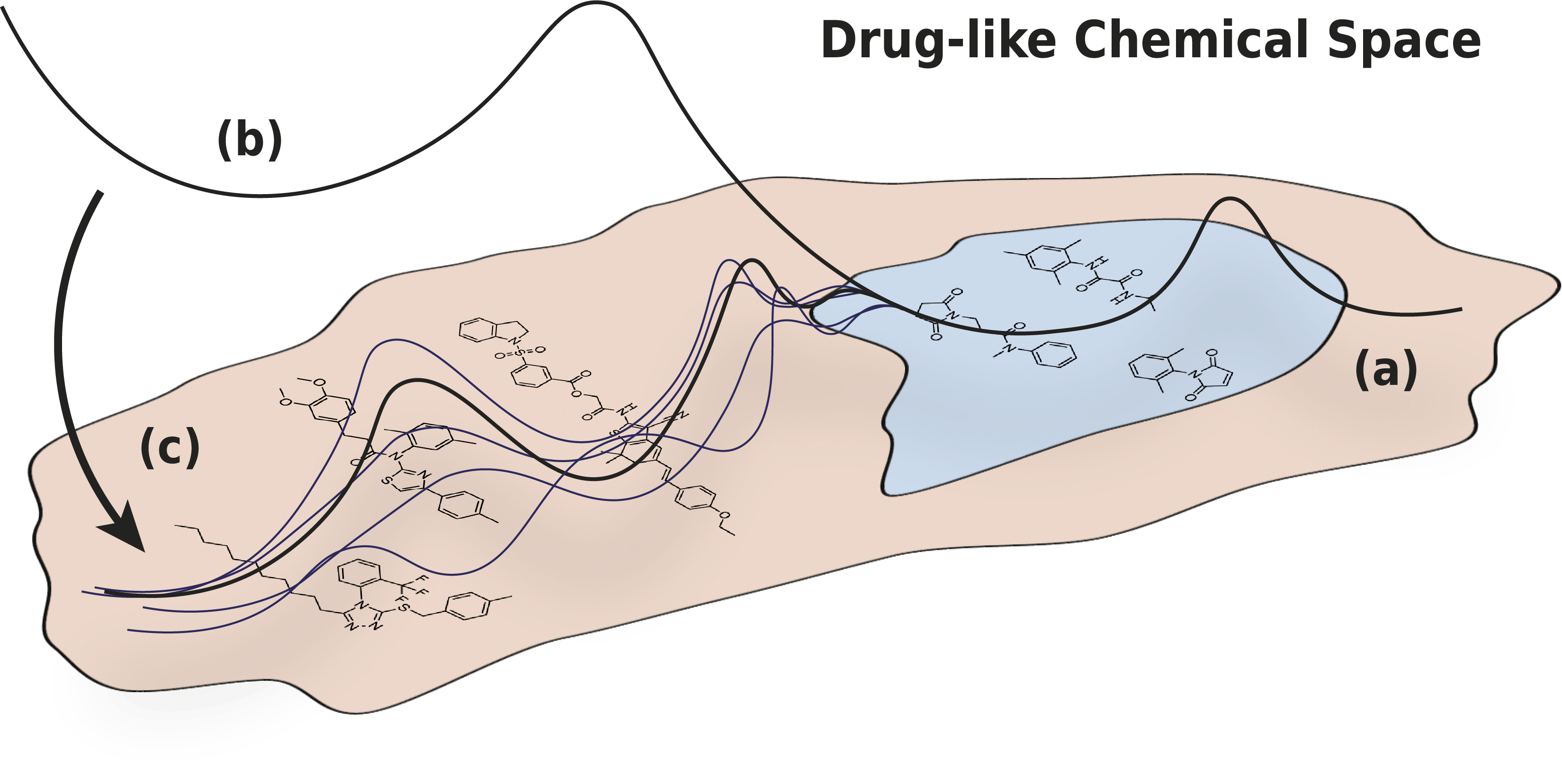}
    \caption{In early-stage drug discovery, bioactivity labels are usually only available for a small and biased subset of compounds (a).
    However, predictions are often most useful for novel molecules that are dissimilar to the ones already explored---an evaluative setting in which deep learning algorithms perform unreliably (b).\\
    We demonstrate that using a regularizing distribution over functions to encode prior knowledge of drug-like chemical space into the modeling process improves the predictive performance and calibration of neural networks in this extrapolative regime (c).}
\label{fig:graphical_abstract}
\vspace*{-5pt}
\end{figure}

A key feature of practical early-stage drug discovery research is the application of predictive models to novel compounds that are \textit{structurally or functionally dissimilar} to molecules that have already been explored (see~\Cref{fig:graphical_abstract}). 
In such an \textit{extrapolative regime}, the practical utility of machine learning systems hinges on their ability to \begin{enumerate*}[label=(\alph*)] \item robustly generalize to unexplored areas of chemical space and \item reliably indicate when they fail to do so by generating well-calibrated predictive uncertainty estimates\end{enumerate*}.
\begin{figure*}[!t]
\centering
\raisebox{0.55in}{\rotatebox{90}{Bioactivity}}
\includegraphics[width=0.95\textwidth]{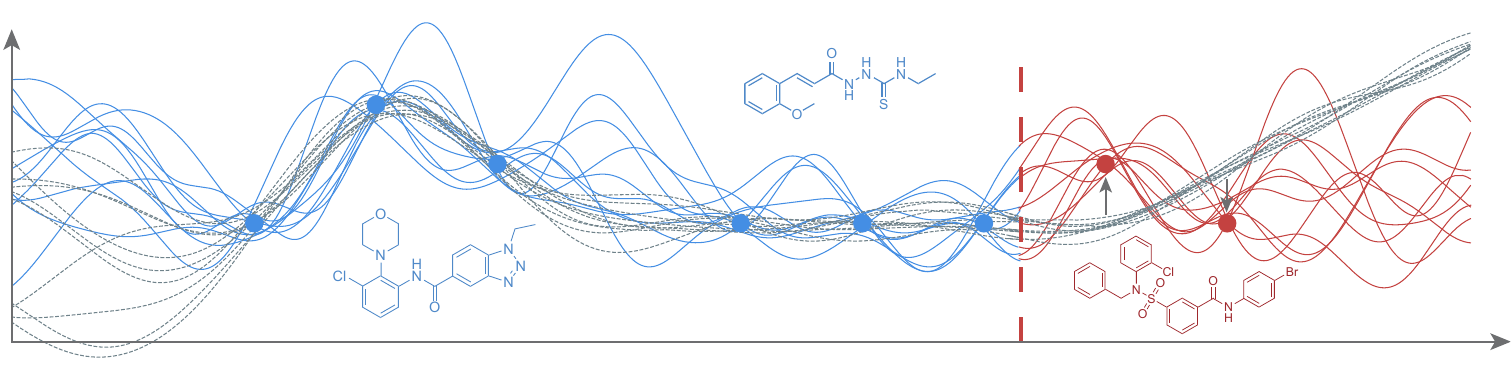}
\put (-335,5) {\color{black}Interpolation}
\put (-105,5) {\color{black}Extrapolation}
\vspace{-10pt}
\caption{When trained on a small and highly biased subset of chemical space, standard neural networks {\color[HTML]{6f828a}(\textbf{gray})} rarely generalize well in the extrapolative regime.
Our approach enables the construction of a problem-informed regularizing prior distribution over functions to place soft constraints on a neural network's hypothesis space, enabling better generalization and uncertainty quantification under covariate shift.
In-distribution training points are shown in {\color[HTML]{448ee4}(\textbf{blue})} and out-of-distribution test points are shown in {\color[HTML]{c44240}(\textbf{red})}.}
\label{fig:science_background}
\vspace*{-5pt}
\end{figure*}
However, standard deep learning algorithms often perform poorly under covariate shift, generating both incorrect and highly miscalibrated predictions \citep{ovadia2019uncertainty, koh2021wilds}. 
This is particularly problematic in the context of early-stage drug discovery, where experimental labels are expensive to acquire and therefore only available for a small and often highly biased subset of compounds.

To improve the predictive performance of deep learning algorithms in such resource-constrained, low-data settings, we may wish to use relevant prior knowledge about the problem domain to specify inductive biases that make some predictive functions more likely than others.
Common approaches to imbuing neural networks with useful inductive biases include \begin{enumerate*}[label=(\alph*)]
    \item pre-training them on larger, potentially unlabeled datasets \citep{finn2017model, tan2018survey, bommasani2021opportunities} and 
    \item adjusting their architectures to mirror appropriate invariances of their input domain \citep{bronstein2017geometric, satorras2021n}.
\end{enumerate*}
However, these approaches are only an indirect---and often insufficiently precise---way of translating explicit modeling preferences into constraints over a neural network's hypothesis space.

In this paper, we present an alternative approach.
To encode domain-informed prior knowledge of the data-generating process into neural network training, we specify a prior distribution over the space of \textbf{Q}uantitative \textbf{S}tructure-\textbf{A}ctivity mappings evaluated at a carefully selected set of context points, and perform \textbf{V}ariational \textbf{I}nference in the resulting probabilistic model (see~\Cref{fig:science_background}). 
We will refer to this method as \textbf{\textsc{q-savi}}.

To demonstrate the practical utility of this approach, we construct a robust evaluation setup based on a carefully pre-processed bioactivity dataset.
We then apply several different techniques to induce strong covariate and label shifts, resulting in challenging and practically meaningful train-test splits.
Finally, we use \textsc{q-savi} to specify explicit and problem-informed prior knowledge of drug-like chemical space and show that this substantially improves the predictive accuracy and calibration of deep learning algorithms in an out-of-distribution setting, outperforming a range of strong self-supervised pre-training, domain adaptation, and ensembling techniques.
\codebox{Code and datasets are provided at}{\begin{center}
    \footnotesize
    \url{https://github.com/leojklarner/Q-SAVI}.
\end{center}}
\vspace*{-10pt}

\newpage

\section{Predicting Properties to Discover Drugs}
\label{sec:background}

The overarching objective of small molecule drug discovery is to identify compounds that modulate a biological target of interest and elicit a therapeutically beneficial response.
Unfortunately, the process of discovering a promising candidate to take into clinical trials is difficult and often unsuccessful, as the search space of viable drug-like molecules \mbox{$\mathcal{X}=\{m_1,m_2, ...\}$} is vast, with estimates of $\vert\mathcal{X}\vert$ ranging from $10^{20}$ to $10^{60}$ \cite{bohacek1996art, ertl2003cheminformatics, polishchuk2013estimation}.
This is compounded by the inherent experimental limitations of medicinal chemistry, meaning that labels can only be acquired for a vanishingly small subset of compounds \mbox{$\calX'\subset\calX$}, with \mbox{$|\calX'|\ll|\calX|$}.
Naturally, this has generated substantial interest in training supervised machine learning algorithms on available data \mbox{$\calD = \{ (\bx_{i}, \by_{i}) | \bx_i\in \calX', \by_i\in\calY\}_{i=1}^N $} to predict the properties of compounds in \mbox{$\calX\setminus\calX'$}.

As the purpose of such models is to accelerate the discovery of novel and more effective therapeutics, predictions are usually desired most on compounds that are meaningfully dissimilar to molecules in $\calX'$.
Were $\calX'$ sampled uniformly from $\calX$, that is, \mbox{$\calX'\sim \mathcal{U}(\calX)$}, these predictions would be made in an interpolative regime, in which standard regularization techniques such as weight decay, dropout \cite{srivastava2014dropout}, and batch normalization \cite{ioffe2015batch} constitute effective approaches to minimizing the expected loss on new samples from $\mathcal{U}(\calX)$.

In practice, however, the composition of $\calX'$ is largely determined by empirical considerations such as compound availability and the preferences and intuitions of medicinal chemists, resulting in a highly biased subsample $\calX'\sim \Tilde{p}_{\calX}$.
This means that, in order to reliably predict the properties of novel and scientifically interesting compounds, it is essential for machine learning algorithms to perform well in an extrapolative regime.
As this requirement is distinct from in-distribution generalization, standard approaches to regularization are unlikely to be effective.

Instead, we propose an alternative regularization scheme---\textsc{Q-SAVI}---that builds on the fact that we are able to approximately sample from $\mathcal{U}(\calX)$ through large chemical databases such as \textsc{ZINC} \citep{irwin2020zinc20} or \textsc{GDB} \cite{polishchuk2013estimation} to specify arbitrary modeling preferences on \mbox{$\calX\setminus\calX'$}.
Specifically, we construct a probabilistic model of neural network \textit{functions} and define a tractable prior distribution over parametric function mappings evaluated at points in $\mathcal{U}(\calX)$. 
We then extend this probabilistic model to include a label-space prior over $\calX$, which encodes contextualized information on $\Tilde{p}_{\calX}$ and $\calY$, and demonstrate empirically that variational inference in this probabilistic model results in neural networks that make accurate predictions in regions of chemical space that they can reliably extrapolate to while generating well-calibrated predictive uncertainty estimates that indicate when correct predictions are unlikely.

\vspace*{-3pt}
\section{Related Work}
\label{sec:related_work}

Starting with foundational attempts to link the electronic properties of different substituents to the reactivity \citep{hammett1937effect} and bioactivity \citep{hansch1962correlation} of benzoic acid derivatives, the problem of predicting the properties of a molecule from its structure has long received considerable attention \citep{cherkasov2014QSAR}.
While simpler algorithms such as support vector machines \citep{cortes1995support} and random forests \citep{breiman2001random} remain a popular choice for such quantitative structure-activity relationship (\textsc{QSAR}) models, recent years have seen substantial interest in applying modern deep learning algorithms to this task \citep{ma2015deep,gawehn2016deep,zhang2017machine}, including important attempts to improve their performance in low-data and out-of-distribution regimes.

\vspace*{-3pt}
\paragraph{Self-supervised pre-training techniques.}
To this end, \citet{hu2019strategies} and \citet{rong2020self} have introduced a range of self-supervised objectives to pre-train graph neural networks and graph transformers on a set of unlabeled molecular structures to generate initializations that can be efficiently fine-tuned on downstream tasks.
However, the out-of-distribution generalization of their approaches was only assessed on scaffold splits---a setting that may underestimate of covariate and label shift encountered in many practical applications \citep{wallach2018most}.

\vspace*{-3pt}
\paragraph{Domain adaptation techniques.}
Building on the fact that biases in the data collection process are often known at training time, domain adaptation and generalization techniques \citep{ganin2016domain,sun2016deep,sagawa2019distributionally,arjovsky2019invariant} aim to improve the performance of deep learning algorithms in out-of-distribution settings by leveraging pre-specified domain indicators. However, these methods---originally developed for image data---have been found to provide limited benefits in the context of molecular property prediction~\citep{ji2022drugood}.

\paragraph{Bayesian inference-based techniques.}
Bayesian Neural Networks (\textsc{BNNs};~\citet{neal1996Bayesian}) provide a principled probabilistic framework for posterior inference over neural networks parameters and have long been explored in the context of drug discovery~\citep{burden1999robust,burden2000use}.
Even though they conceptually guarantee robustness in low-data regimes, their empirical performance often falls short of ensembling techniques or even standard stochastic gradient descent~\citep{ovadia2019uncertainty, foong2019inbetween, farquhar2020radial}, including in the context of molecular property prediction~\citep{ryu2019bayesian,zhang2019bayesian}.

While these approaches may improve the robustness of deep learning algorithms in some settings, they are limited in the extent to which they can encode problem-specific modeling preferences that, for example, encourage high predictive uncertainty away from the training data or specify prior knowledge of synthetic accessibility and patentability.
For instance, the standard parameter-space formulation of \bnns precludes the specification of semantically meaningful prior information due to the highly non-linear and complex relationship between a neural network's parameters and the functions they encode.

Building on recent work that aims to address the shortcomings of \bnns (e.g., in specifying meaningful prior distributions and providing reliable uncertainty quantification) via function-space variational inference~\citep{sun2019fbnn, rudner2021tractable, Rudner2022sfsvi}, we reframe \textsc{QSAR} modeling as inferring a posterior distribution over functions.
We do so by specifying a prior distribution over function mappings along with a prior distribution over function evaluation points and performing variational inference in this probabilistic model, which allows us to explicitly encode prior beliefs about the distribution over functions as well as about the structure of the input space into neural network training.

\vspace*{-3pt}
\section{Quantitative Structure-Activity VI}
\label{sec:method}

Consider the supervised learning setup outlined in~\Cref{sec:background}, with the objective of training a machine learning model on the experimental labels of $N$ independent and identically distributed samples drawn from a biased subset of chemical space, resulting in the data realizations \mbox{${\calD = \{\bx_i, \by_i\}_{i=1}^N} = (\bx_\calD, \by_\calD)$} of inputs \mbox{$\bx_i \in \calX'\subset\calX$} and labels \mbox{$\by_i \in \calY$}, where \mbox{$\calY \subseteq \real^K$} for regression and \mbox{$\calY \subseteq \{0, 1\}^K$} for classification tasks with $K$ labels.

Let $p_{\bY | f(\bX ; \bTheta)}$ be an observation model of the labels $\bY$ given a latent stochastic function $f(\bX ; \bTheta):\calX\times\R^P\to\calY$ induced by a set of stochastic parameters $\bTheta\in\R^P$ and evaluated at a set of input points $\bX\in\calX$. Additionally, let $p_{f(\bX ; \bTheta)}$ be a prior distribution over such latent stochastic functions.
$p_{\bY | f(\bX ; \bTheta)}(\by_{\calD} \vbar f(\bx_{\calD} ; \btheta))$ is then the likelihood of observing labels $\by_{\calD}$ under $f(\bx_{\calD} ; \btheta)$ --- a realization of the stochastic function evaluated at inputs $\bx_{\calD}$.

Instead of formulating the posterior inference problem as finding the posterior distribution over stochastic parameters $\bTheta$, we follow \citet{rudner2021tractable} and reframe variational inference in stochastic neural networks as finding a posterior distribution over the \textit{latent stochastic functions} $f(\bx_{\calD}; \bTheta)$ at the training points $\bx_{\calD}$.
In particular, while the parameter-space Bayesian inference problem is given by
\begin{align}
\label{eq:bayes_ps}
    p_{\bTheta | \calD}(\bTheta \vbar \calD)
    =
    \frac{p_{\bY | \bTheta, \bX}(\by_{\calD} \vbar \btheta, \bx_{\calD}) \, p_{\bTheta}(\btheta)}{p_{\bY | \bX}(\by_{\calD} \vbar \bx_{\calD})} ,
\end{align}
the inference problem over $f(\bx_{\calD} ; \btheta)$ is expressed by
\begin{align}
\begin{split}
\label{eq:bayes_fs}
    &
    p_{f(\bX ; \bTheta) | \calD}(f(\bx_{\calD} ; \btheta) \vbar \calD)
    \\
    &
    =
    \frac{p_{\bY | f(\bX ; \bTheta)}(\by_{\calD} \vbar f(\bx_{\calD} ; \btheta)) \, p_{f(\bX ; \bTheta)}(f(\bx_{\calD} ; \btheta))}{p_{\bY | \bX}(\by_{\calD} \vbar \bx_{\calD})},
\end{split}
\end{align}
which includes an explicit dependence on the function-space prior evaluated at the training points $p_{f(\bX ; \bTheta)}(f(\bx_{\calD} ; \btheta))$, which allows us to specify arbitrary preferences for suitable parametric function mappings $f$.

To show how the inference problem in~\Cref{eq:bayes_ps} and~\Cref{eq:bayes_fs} are related, note that for a prior distribution over parameters $p_{\bTheta}$, the prior distribution $p_{f(\bX ; \btheta)}$ over $f(\bx_{\calD} ; \btheta)$ induced by $p_{\bTheta}$ is given by
\begin{align}
\begin{split}
    &
    p_{f(\bX ; \bTheta)}(f(\bx_{\calD} ; \btheta))
    \\
    &
    =
    \int_{\R^{P}} p_{\bTheta}(\btheta') \, \delta( f(\bx_{\calD} ; \btheta) - f(\bx_{\calD} ; \btheta') ) \dee \btheta' ,
\end{split}
\end{align}
and, similarly, the posterior distribution $p_{f(\bX ; \bTheta) | \calD}$ over $f(\bx_{\calD} ; \btheta)$ induced by the posterior distribution over parameters $p_{\bTheta | \calD}$ is given by
\begin{align}
\begin{split}
    &
    p_{f(\bX ; \bTheta) | \calD}(f(\bx_{\calD} ; \btheta) \vbar \calD)
    \\
    &
    =
    \int_{\R^{P}} p_{\bTheta | \calD}(\btheta' \vbar \calD) \, \delta( f(\bx_{\calD} ; \btheta) - f(\bx_{\calD} ; \btheta') ) \dee \btheta' ,
\end{split}
\end{align}
where $\delta(\cdot)$ is the Dirac delta function~\citep{wolpert1993fsmap,Rudner2022fsvi}.
In the remainder of this section, subscripts will be dropped from probability density functions when the dependence is clear from context.

We will now extend this function-space formulation of Bayesian inference to define a probabilistic model that is able to integrate prior knowledge of the full input space $\calX$ beyond a biased subset of training points $\bx_\calD\subseteq\calX'$.
Specifically, we extend the probabilistic model above to the random variables $f(\{ \bX, \bX_{\calC} \}; \bTheta)$ and $\bX_{\calC}$, where $\bX_{\calC}\subseteq\calX\setminus\calX'$ is a set of \textit{context points}, yielding the posterior distribution
\begin{align}
\label{eq:bayes_fs_context}
    &
    p(f(\{ \bx_{\calD}, \bx_{\calC} \} ; \btheta), \bx_{\calC} \vbar \calD)
    \\
    &
    =
    \frac{p(\by_{\calD} \vbar f(\bx_{\calD} ; \btheta)) \, p(f( \{ \bx_{\calD}, \bx_{\calC} \} ; \btheta) \vbar \bx_{\calD}, \bx_{\calC} ) \, p(\bx_{\calC}) }{p(\by_{\calD} \vbar \bx_{\calD})}
    \nonumber
\end{align}
where, for a stochastic function evaluation $f(\{ \bx_{\calD}, \bx_{\calC} \} ; \btheta)$ defined by a valid stochastic process over $f(\cdot \,; \bTheta)$, the likelihood $p(\by_{\calD} \vbar f(\bx_{\calD} ; \btheta))$ and marginal likelihood $p_{\bY | \bX}(\by_{\calD} \vbar \bx_{\calD})$ are independent of $f( \bX_{\calC} ; \bTheta)$ and $\bX_{\calC}$ by marginal consistency.

For non-linear function mappings \mbox{$f : \calX \times \R^P \rightarrow \calY$} parameterized by high-dimensional $\bTheta \in \R^P$, the inference problem specified in~\Cref{eq:bayes_fs_context} is analytically intractable.
Instead, we may frame it variationally as
\begin{align}
    \min_{q_{\bTheta} \in \calQ_{q_{\bTheta}}} \DD_{\textrm{KL}}( q_{f(\{ \bX, \bX_{\calC} \} ; \bTheta), \bX_{\calC}} \,\|\, p_{f(\{ \bX, \bX_{\calC} \} ; \bTheta), \bX_{\calC} | \calD} )
\end{align}
for some variational distribution over parameters $q_{\bTheta}$ in a variational family $\calQ_{q_{\bTheta}}$~\citep{wainwright2008vi}.
Letting the variational distribution factorize as
\begin{align}
    q_{f(\{ \bX, \bX_{\calC} \} ; \bTheta), \bX_{\calC}}
    \defines
    q_{f(\{ \bX, \bX_{\calC} \} ; \bTheta)} q_{\bX_{\calC}}
    =
    q_{f(\{ \bX, \bX_{\calC} \} ; \bTheta)} p_{\bX_{\calC}} ,
\end{align}
and assuming that $q_{\bX_{\calC}} = p_{\bX_{\calC}}$, we can reformulate the inference problem above in a simplified form as
\begin{align}
    \min_{q_{\bTheta} \in \calQ_{q_{\bTheta}}} \E_{p_{\bX_{\calC}}} \left[ \DD_{\textrm{KL}}( q_{f(\bX ; \bTheta) | \bX_{\calC}} \,\|\, p_{f(\bX ; \bTheta) | \bX_{\calC}, \calD} ) \right] ,
\end{align}
which can in turn be equivalently expressed as
\begin{align}
\label{eq:objective}
\begin{split}    
    &
    \max_{q_{\bTheta} \in \calQ_{q_{\bTheta}}} \bigg\{ \E_{q_{\bTheta} p_{\bX_{\calC}}} \left[ \log p(\by_{\calD} \vbar f(\bx_{\calD} ; \btheta)) \right]
    \\
    &
    \qquad\qquad~~
    - \DKL{q_{f(\{ \bX, \bX_{\calC} \} ; \bTheta)}}{p_{f(\{ \bX, \bX_{\calC} \} ; \bTheta)}} \bigg\} .
\end{split}
\end{align}
If $p_{\bTheta}$ is chosen to be an isotropic Gaussian distribution and $\calQ_{q_{\bTheta}}$ is the family of mean-field Gaussian distributions, the prior and variational distributions in~\Cref{eq:objective} can be approximated using the local linearization scheme introduced in~\citet{Rudner2022fsvi}.
These approximations result in a factorized variational objective, making stochastic variational inference and stochastic gradient-based optimization techniques applicable~\citep{hinton1993keeping,graves2011practical,hoffman2013svi,blundell2015mfvi}.

By enabling the specification of the context point distribution $p_{\bX_{\calC}}$
and the prior distribution over functions $p_{f(\{ \bX, \bX_{\calC} \} ; \bTheta)}$, this framework enables us to explicitly encode 
arbitrary modeling preferences as distributions that place high probability mass on relevant regions of the input domain and specify prior knowledge of preferred parametric function mappings on unlabelled data points.

After optimizing the variational objective with respect to the parameters of $q_{\bTheta}$, we obtain samples from approximate posterior predictive distribution through
\begin{align}
\begin{split}
    \hspace*{-3pt}q(\by_{\ast} \vbar \bx_{\ast})
    &
    \approx
    \frac{1}{M_{\ast}} \sum\nolimits_{j=1}^{M_{\ast}} p(\by_{\ast} \vbar f(\bx_{\ast} ; \btheta^{(j)})) ,
\end{split}
\end{align}%
with $\btheta^{(j)} \sim q_{\bTheta}$ and $M_{\ast}$ being the number of Monte Carlo samples used to estimate the predictive distribution.

\section{Empirical Evaluation}
\label{sec:experiments}

To demonstrate the practical utility of \textsc{q-savi}, we establish a robust evaluation setup:
In~\Cref{sec:data_curation}, we argue that many commonly-used bioactivity datasets may not be able to meaningfully assess the extrapolative power of supervised machine learning algorithms and present a carefully cleaned and pre-processed alternative dataset and in~\Cref{sec:data_shift}, we define an appropriate set of statistics to quantify covariate and label shifts in chemical space and use them to investigate the extent to which different splitting techniques induce data shift.
In~\Cref{sec:models_and_eval}, we then use this experimental setup to demonstrate that employing \textsc{q-savi} to incorporate domain-informed prior knowledge into the modeling process leads to significant gains in predictive accuracy and calibration, outperforming a range of strong self-supervised pre-training, domain adaptation, and ensembling techniques.
Finally, in~\Cref{sec:merck-datasets}, we show that these strong empirical results extend to real-world production settings by evaluating our method on the time-split data presented in \citet{ma2015deep}.

\subsection{Curating an Appropriate Dataset}
\label{sec:data_curation}

A fundamental obstacle to training and evaluating \textsc{QSAR} models in the public domain is the scarcity of sufficiently large datasets with high-quality labels~\citep{schneider2020rethinking}.
Even though collections of publicly available bioactivity data exist~\citep{wu2018moleculenet, huang2021therapeutics}, they are often sourced directly from repositories of high-throughput screening (HTS) data such as \textsc{PubChem}~\citep{kim2019pubchem}, \textsc{ChEMBL}~\citep{mendez2019chembl} or \textsc{ToxCast}~\citep{richard2016toxcast} without significant filtering or pre-processing.
While this approach maximizes the number of available data points, it may reduce the discriminative power of model performance comparisons.
For instance, a well-known problem of confirmatory dose-response screens---which make up the bulk of measurements in the above repositories---is that they usually contain a large number of reproducible false positive readouts (in many cases up to 95\% of hits~\citep{thorne2010apparent}) caused by molecular substructures that interfere with an assay's readout system \citep{baell2010new,dahlin2015pains}.
Using such data without further processing runs the risk of simply testing for the ability of algorithms to memorize these substructures instead of assessing meaningful extrapolative performance \citep{klarner2022bias}.

To curate a dataset of sufficient quality to enable an informative comparison of predictive models, we used the measurement meta-data of bioactivity and toxicity screens 
to prioritize certain data points for further inspection.
After surveying the publications associated with the most promising datasets, we selected a high-quality screening campaign for inhibitors of the development of liver-stage malaria parasites for further processing \citep{antonova2018open}.

Specifically, we retrieved and reprocessed the raw measurement data to remove likely false positives and other experimental artifacts, yielding a binary classification dataset with 7,301 inactive and 849 active molecules, each measured in biological duplicate and confirmed as a true positive or negative through a set of quality-assuring counter-screens (see~\Cref{appsec:dataset_processing} for full details).

\begin{figure}[t]
    \centering
    \subfloat{{\includegraphics[width=0.48\columnwidth]{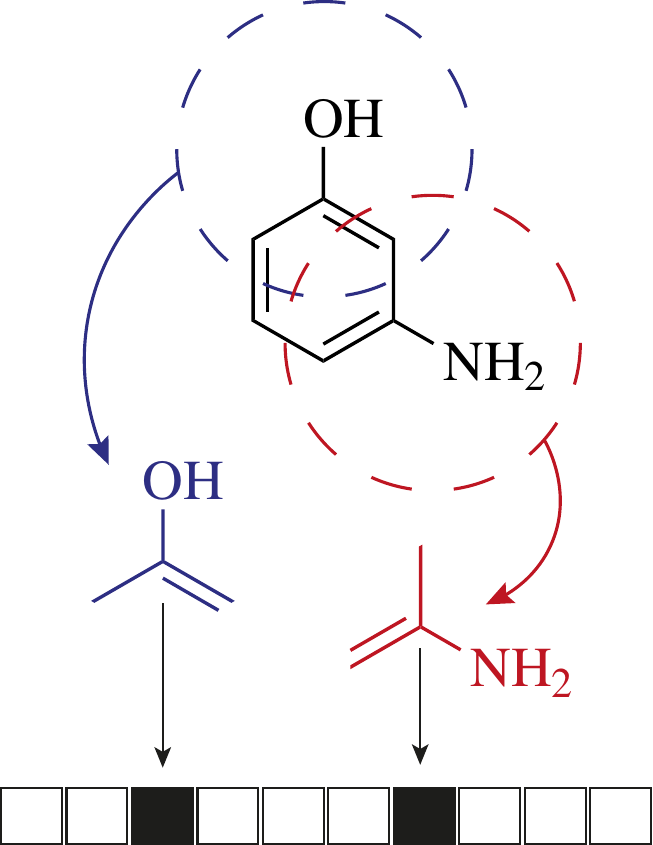}}}
    \hfill
    \subfloat{{\includegraphics[width=0.48\columnwidth]{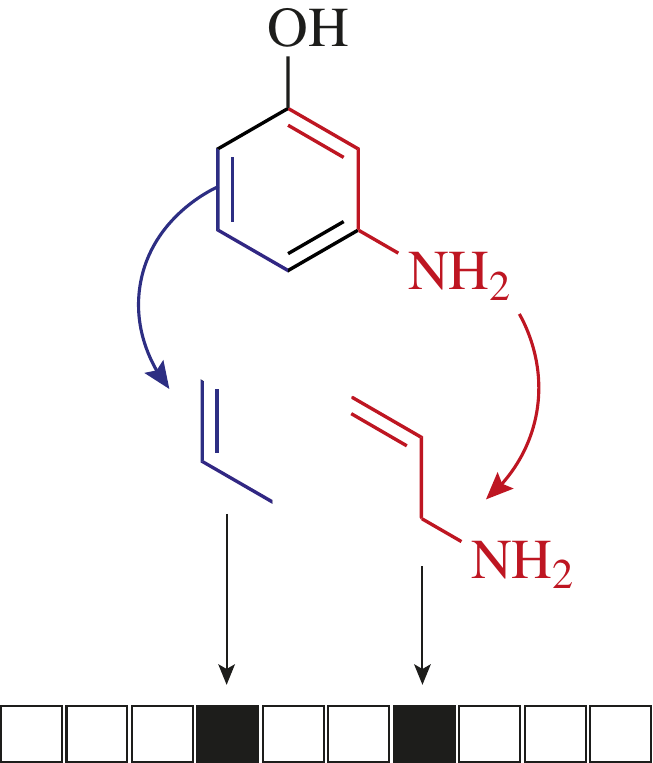}}}
    \caption{Schematic representation of extended-connectivity fingerprints (\textbf{left}) and \textsc{RDKit} fingerprints (\textbf{right}).
    Both methods operate on the topological graph of a molecule and enumerate all labeled subgraphs up to a certain diameter, differing only in the space of substructures they consider.
    While \textsc{rdkitFPs} enumerate subgraphs of any shape, \textsc{ECFPs} are restricted to radial substructures.
    Once extracted, this set of subgraphs is hashed into a fixed-length bit vector.}
    \label{fig:featurizaton}
\end{figure}

\subsection{Inducing and Quantifying Data Shift}
\label{sec:data_shift}

\paragraph{Featurization.}
Commonly used techniques to numerically represent the structural properties of a molecule include strings, graphs, and topological fingerprints. 
For the following experiments, each molecule was featurized as both an extended-connectivity fingerprint (\textsc{ECFP}; \citet{rogers2010extended}) and an \textsc{RDKit} fingerprint (\textsc{rdkitFPs}), using the respective implementations in the open-source cheminformatics package \textsc{RDKit}~\citep{landrum2013rdkit}.
An illustration of this process is presented in~\Cref{fig:featurizaton}.

\paragraph{Statistics for covariate and label shift.}
To evaluate the extent to which different train-test splits induce covariate and label shift, we identified  a set of suitable two-sample test statistics and used it to quantify the dissimilarity of the marginal covariate and label distributions of the respective training and test sets 
$\calD_\text{tr}=(\bX_\text{tr}, \by_\text{tr})$ and $\calD_\text{te}=(\bX_\text{te}, \by_\text{te})$.

Since $\by_\text{tr}$ and $\by_\text{te}$ consist of binary indicators of antimalarial activity, well-established categorical statistics such as Fisher's exact test~\citep{upton1992fisher} are applicable.
In the following, its negative logarithmic p-value is used as a scalar indicator of label shift.  %

Defining a corresponding statistic to quantify covariate shift between two sets of molecules is more challenging, as they constitute disjoint sets of discrete objects.
For this purpose, we used the maximum mean discrepancy (\textsc{MMD}) metric~\citep{gretton2012kernel} to quantify the difference between two samples of molecules as the distance between the embeddings of their expectations in a reproducing kernel Hilbert space (\textsc{RKHS}) defined by some mapping $\phi:\calX\to\mathcal{H}$ and an associated kernel function $k(\bx_i, \bx_j)=\langle\phi(\bx_i), \phi(\bx_j)\rangle_\mathcal{H}$.
An empirical estimator of this statistic is obtained by
\begin{align*} 
    &
    \operatorname{\textsc{MMD}}^2(\bX_\text{tr},\bX_\text{te})
    \\
    &
    =
    \left\lVert \mathbb{E}_{\bx\in\bX_\text{tr}}\left[\phi(\bx)\right] - \mathbb{E}_{\bx\in\bX_\text{te}}\left[\phi(\bx)\right] \right\rVert_\mathcal{H}^2
    \\
    &
    =
    \mathbb{E}_{\bx_i, \bx_j\in\bX_\text{tr}}\left[k(\bx_i, \bx_j)\right] + \mathbb{E}_{\bx_i, \bx_j\in\bX_\text{te}}\left[k(\bx_i, \bx_j)\right]
    \\
    &
    \qquad
    - 2\mathbb{E}_{\bx_i\in\bX_\text{tr}, \bx_j\in\bX_\text{te}}\left[k(\bx_i, \bx_j)\right],
\end{align*}
using the Jaccard/Tanimoto similarity coefficient 
\begin{align*}
k_\text{jac}(\bx_i,\bx_j)=\frac{\bx_i\cap\bx_j}{\bx_i\cup\bx_j}=\frac{\left\langle\bx_i,\bx_j\right\rangle}{\left\lVert\bx_i\right\rVert^2+\left\lVert\bx_j\right\rVert^2-\left\langle\bx_i,\bx_j\right\rangle}
\end{align*}
as an appropriate similarity metric, both due to its established use in the cheminformatics community~\citep{bajusz2015tanimoto} and the favorable properties of the \textsc{RKHS} that it induces.
The \textsc{MMD} statistic is only valid if the mean embedding $\mathbb{E}_{\bx\in\bX}\left[\phi(\bx)\right]$ is injective, which is the case for strictly positive definite kernels operating in discrete domains~\citep{borgwardt2006integrating}, such as $k_\text{jac}$~\citep{bouchard2013proof}.

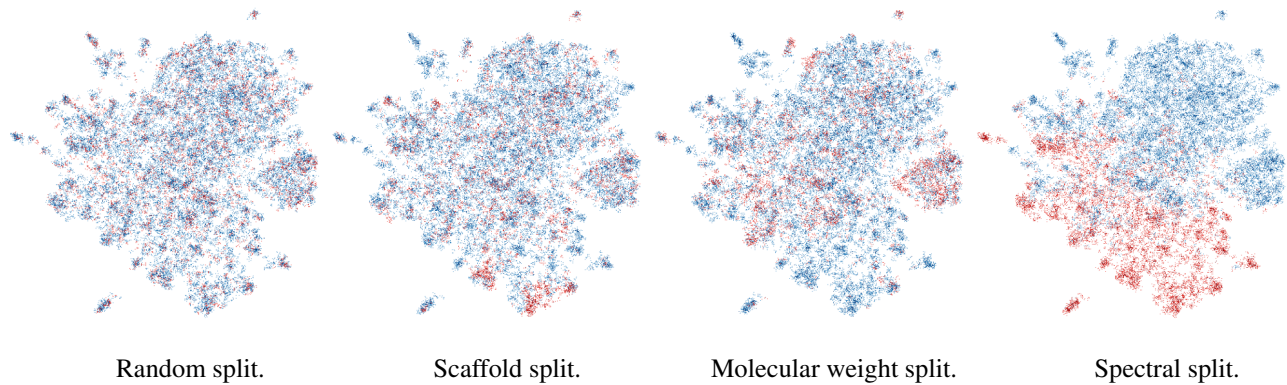
\begin{figure*}
    \centering
    \includegraphics[width=\textwidth]{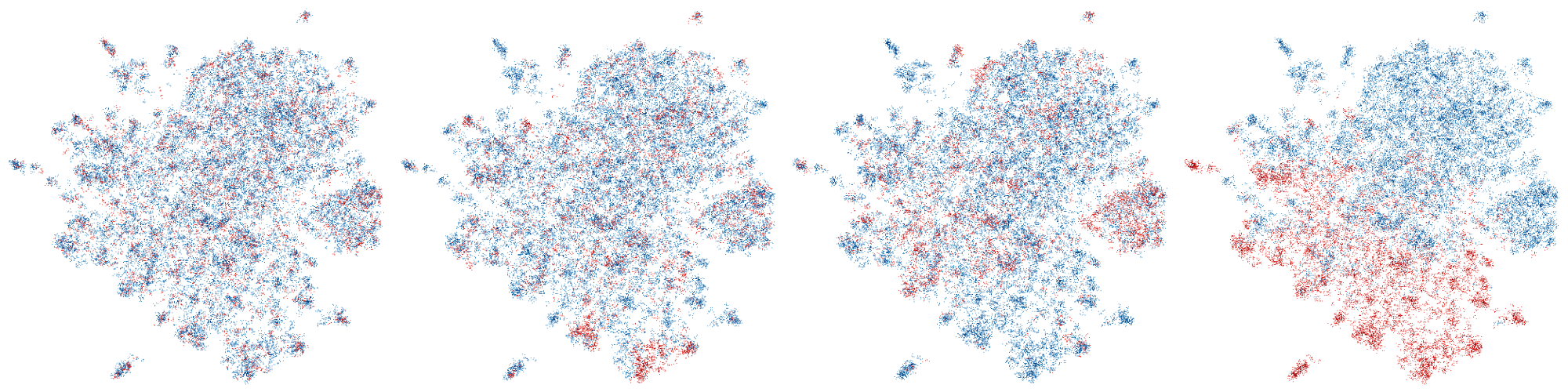}
    \put(-445, -20){Random split.}
    \put(-325, -20){Scaffold split.}
    \put(-220, -20){Molecular weight split.}
    \put(-75, -20){Spectral split.}
    \caption{A visual comparison of the covariate shift induced by different train-test splits using rdkit fingerprints, colored in {\color[HTML]{448ee4}\textbf{blue}} and {\color[HTML]{c44240}\textbf{red}} respectively. While random and scaffold splits lead to relatively similar training and test sets, molecular weight and spectral splits induce significantly stronger covariate shifts. The plots were generated using UMAP dimensionality reduction \citep{mcinnes2018umap}.}
\label{fig:split_umap}
\end{figure*}

\begin{figure}[H]
    \centering
    \vspace{0.5em}
    \includegraphics[width=0.8\columnwidth]{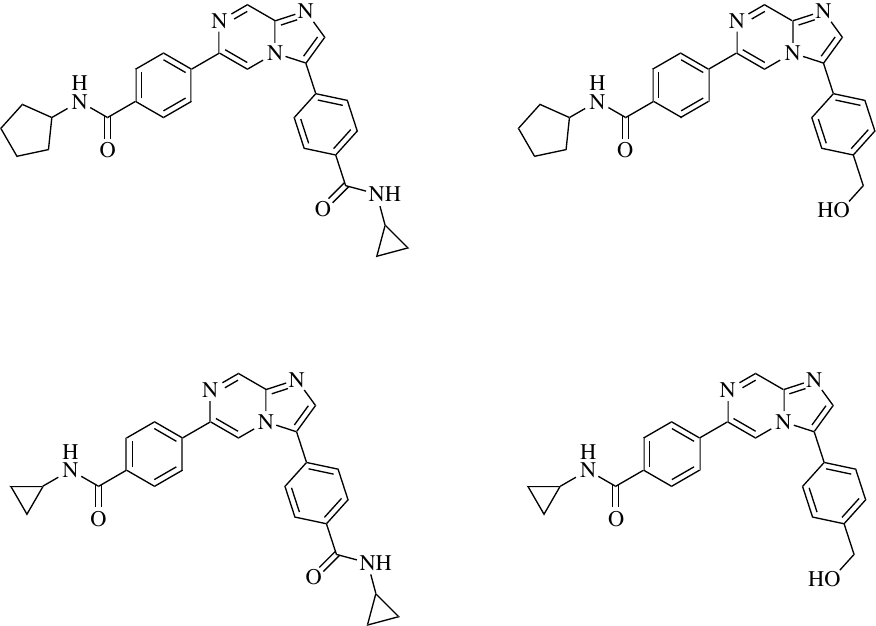}
    \caption{Even molecules with nominally different Bemis-Murcko scaffolds can exhibit a high degree of structural and functional similarity. Depicted are four structurally similar molecules from our antimalarial dataset that are assigned to four different scaffolds.}
    \label{fig:scaffolds}
\end{figure}

\paragraph{Random and scaffold splits.}
\label{par:data_splits}
Equipped with the appropriate statistical tools to quantify distributional similarities, we investigated the extent to which different train-test splits are able to emulate practically relevant covariate and label shifts, beginning with the two most popular approaches of splitting data either randomly or by scaffold. 
While \emph{random splits} are commonly used in many domains, they are known to produce unrealistically optimistic performance estimates in the context of molecular property prediction.
This is a consequence of the biased composition of many experimental datasets, which often contain structurally similar compounds from so-called chemical series.
As these often exhibit very similar properties, distributing them evenly across data splits leads to a de-facto overlap between training and test sets that incentivizes overfitting and memorization~\citep{wallach2018most}.
\emph{Scaffold splits} attempt to mitigate this shortcoming by mapping each molecule to an overarching compound class---usually its Bemis-Murcko scaffold~\citep{bemis1996properties, bemis1999properties}---and splitting the data so that all molecules of a given scaffold are assigned to the same partition.
However, this approach often results in a similar pathology, as even molecules with nominally different scaffolds can exhibit a high degree of structural and functional similarity, as illustrated in~\Cref{fig:scaffolds}.

\vspace{-1em}
\paragraph{Molecular weight and spectral splits.}
To facilitate the comparison of models in an extrapolative regime, we explored two alternative approaches.
A straightforward \textit{molecular weight split} was used to induce data shift by assigning molecules into training and test sets based on a molecular weight cut-off, relying on the correlation of molecular size and binding strength to also induce strong label shift \citep{hopkins2014role}.
More rigorously, we developed a clustering-based \emph{spectral split} to generate data splits that are guaranteed to exhibit maximal covariate shift under the \textsc{MMD} statistic.
By interpreting the Jaccard kernel Gram matrix $\mathbf{W}^\text{jac}\in[0,1]^{\vert\bX_\calD\vert\times\vert\bX_\calD\vert}$ of a given set of molecules $\bX_\calD$ as the weighted adjacency matrix of a fully-connected similarity graph $S$, well-established spectral clustering algorithms~\citep{von2007tutorial} can be employed to identify an optimal partitioning of $S$ that maximizes the similarity within and minimizes the similarity between partitions.

We present a comparison of the resulting covariate and label shift statistics in~\Cref{tab:data_shift}, which shows that molecular weight and spectral clustering-based splits generate a significantly more extrapolative evaluation setup than random and scaffold splits. This is substantiated by the qualitative visualization presented in~\Cref{fig:split_umap}.
\vspace{-0.5em}
\begin{table}[H]
\centering
\caption{
    A summary of the covariate and label shifts induced by the different train-test splits presented in~\Cref{sec:data_shift}, using rdkit and extended-connectivity (EC) fingerprints. 
    Covariate shift is quantified as the Jaccard kernel-based {\textsc{MMD}} statistic, while label shift is quantified as the negative $\log p$-value of Fisher's exact test.
}
\label{tab:data_shift}
\vspace{1pt}
\small
\begin{tabularx}{\columnwidth}{lcc}
\toprule
Split & Covariate Shift \emph{(rdkit, EC)} & Label Shift \emph{(rdkit, EC)}\\
\midrule
Random & $0.00$, $0.00$ & \num{0.00}\\
Scaffold & $0.08$, $0.07$ & \num{4.23} \\
Weight & $0.14$, $0.10$ & $\mathbf{61.96}$ \\
Spectral & $\mathbf{0.34}$, $\mathbf{0.25}$ &  $17.49$, $50.05$ \\
\bottomrule
\end{tabularx}
\end{table}

\subsection{Model Construction, Baselines \&  Results}
\label{sec:models_and_eval}

\paragraph{Model construction.}
Using the increasingly data-shifted splits constructed in \Cref{sec:data_curation,sec:data_shift}, we assessed \textsc{q-savi} with respect to its ability to improve the predictive accuracy and calibration of deep learning algorithms under covariate and label shifts.
By leveraging the option to specify both an arbitrary context point distribution $p_{\bX_{\calC}}$ and a prior distribution over parametric function mappings $p_{f(\{ \bX, \bX_{\calC} \} ; \bTheta)}$, we used \textsc{q-savi} to encode relevant information about both the input domain and the label space of the problem setup into the model.
Specifically, we precomputed the featurizations of a uniform subsample of the \textsc{ZINC} database of commercially available compounds \citep{irwin2020zinc20} and used them to construct a uniform context point distribution $p_{\bX_{\calC}}=\mathcal{U}(\Bar{\calX})$ over a set of \num{2e6} synthetically accessible drug-like molecules $\Bar{\calX}$. 
Additionally, we used the prior distribution $p_{f(\{ \bX, \bX_{\calC} \}; \bTheta)}$ over parametric mappings to encode an informative function-space prior that encourages high predictive uncertainty in unexplored regions of chemical space, counteracting the likelihood term in~\Cref{eq:objective} to generate better predictive uncertainty estimates.

\paragraph{Baselines.}
We compared the performance of the resulting probabilistic model to a range of standard baselines and state-of-the-art pre-training and domain adaptation techniques.
The simplest of these models is regularized \textbf{logistic regression}, which is expected to underperform in an extrapolative regime due to the linearity of its logit function.
While \textbf{random forest classifiers}~\citep{breiman2001random} represent a more flexible baseline with strong in-distribution generalization guarantees, they generally exhibit coarser decision boundaries at the fringes of the training distribution that are unlikely to perform well on covariate-shifted inputs.
Standard deep learning methods such as multi-layer perceptrons \textbf{(MLPs)} have an even higher representational capacity, yet also generally underperform under data shift, yielding both incorrect and highly overconfident predictions~\citep{ovadia2019uncertainty, koh2021wilds}.
\textbf{Deep ensembles} are an effective technique to improve the predictive performance of MLPs by averaging the predictive distributions of a set of independently trained neural networks~\citep{lakshminarayanan2017simple}.
To investigate the extent to which existing self-supervised pre-training techniques and more expressive model architectures impact the performance of deep learning algorithms in this setting, we fine-tuned graph isomorphism networks (\textbf{GINs}; \citet{xu2018powerful}) provided by \citet{hu2019strategies} both from scratch and from initializations that were pre-trained on compounds from the ZINC database using \emph{context prediction} and \emph{attribute masking} objectives.
Additionally, we fine-tuned the graph transformer (\textbf{GROVER}) proposed by \citet{rong2020self} from a pre-trained initialization that was optimized on molecules from the ZINC and ChEMBL databases using self-supervised contextual property and graph-level motif prediction techniques.
Finally, we adapted a range of domain adaptation and generalization techniques, including invariant risk minimization (\textbf{IRM}; \citet{arjovsky2019invariant}), group-distributionally robust training (\textbf{GroupDRO}; \citet{sagawa2019distributionally}), domain-adversarial networks (\textbf{DANN}; \citet{ganin2016domain}), and deep correlation alignment (\textbf{DeepCoral}; \citet{sun2016deep}) from \citet{ji2022drugood} who provided them with data split-specific domain indicators. 

\setlength{\tabcolsep}{5.0pt}
\begin{table*}[t]
\centering
\caption{
    An overview of the test set performance of each model for each data split and featurization technique, quantified by the \textsc{AUC-ROC} ($\uparrow$) and the \textsc{Brier score} ($\downarrow$).
    All entries indicate the mean and standard errors computed over $10$ independent training runs with different random seeds.
    The best models within a margin of statistical significance are highlighted in bold.
}
\small
\label{tab:results}
\vspace{1pt}
\begin{adjustbox}{max width=\textwidth}
\begin{tabular}{cl|cc|cc|cc|cc}
\toprule
 \multicolumn{2}{c|}{\multirow{2}{*}{Model \& Featurization}} & \multicolumn{2}{c|}{Spectral Split} & \multicolumn{2}{c|}{Weight Split} & \multicolumn{2}{c|}{Scaffold Split} & \multicolumn{2}{c}{Random Split}\\
 & & ECFP & rdkitFP &  ECFP & rdkitFP&  ECFP  & rdkitFP&  ECFP  & rdkitFP \\
\bottomrule
\multicolumn{9}{c}{}\\
\toprule
\parbox[t]{1.5mm}{\multirow{10}{*}{\rotatebox[origin=c]{90}{\textsc{AUC-ROC} ($\uparrow$)}}} & Logistic Regression & $.583\pms{.000}$ & $.551\pms{.000}$ & $.626\pms{.000}$ & $.632\pms{.000}$ & $.684\pms{.000}$ & $.698\pms{.000}$ & $.704\pms{.000}$ & $.687\pms{.000}$\\
& Random Forest & $.576\pms{.009}$ & $.552\pms{.006}$ & $.592\pms{.006}$ & $.567\pms{.004}$ & $.605\pms{.004}$ & $.642\pms{.003}$ & $.696\pms{
.002}$ & $.690\pms{.002}$\\
& MLP & $.574\pms{.006}$ & $.571\pms{.003}$ & $.614\pms{.004}$ & $.577\pms{.005}$ & $.625\pms{.010}$ & $.631\pms{.014}$ & \cellcolor[gray]{0.9}$\mathbf{.720\pms{.002}}$ & $.692\pms{.004}$\\
& Deep Ensemble & $.589\pms{.006}$ & $.571\pms{
.002}$ & $.644\pms{.001}$ & $.594\pms{.002}$ & $.679\pms{.001}$ & \cellcolor[gray]{0.9} $\mathbf{.697\pms{.003}}$ & \cellcolor[gray]{0.9}$\mathbf{.720\pms{.001}}$ & \cellcolor[gray]{0.9} $\mathbf{.710\pms{.003}}$\\
\cmidrule(l){2-10}
& GIN & $.549\pms{.009}$ & $.551\pms{.007}$ & \multicolumn{2}{c|}{$.582\pms{.007}$} & \multicolumn{2}{c|}{$.664\pms{.005}$} & \multicolumn{2}{c}{$.685\pms{.004}$}\\
& GIN (attr masking) & $.588\pms{.004}$ & $.559\pms{.010}$ & \multicolumn{2}{c|}{$.625\pms{.004}$} & \multicolumn{2}{c|}{\cellcolor[gray]{0.9}$\mathbf{.700\pms{.002}}$} & \multicolumn{2}{c}{$.705\pms{.002}$}\\
& GIN (context pred) & $.541\pms{.005}$ & $.566\pms{.009}$ & \multicolumn{2}{c|}{$.621\pms{.003}$} & \multicolumn{2}{c|}{$.674\pms{.003}$} & \multicolumn{2}{c}{\cellcolor[gray]{0.9} $\mathbf{.713\pms{.003}}$}\\
& Grover & $.574\pms{.002}$ & $.544\pms{.006}$ & \multicolumn{2}{c|}{$.623\pms{.003}$} & \multicolumn{2}{c|}{$.689\pms{.003}$} & \multicolumn{2}{c}{$.701\pms{.001}$}\\
\cmidrule(l){2-10}
& \textbf{Q-SAVI} & \cellcolor[gray]{0.9}$\mathbf{.606\pms{.003}}$ & \cellcolor[gray]{0.9}$\mathbf{.603\pms{.006}}$ & \cellcolor[gray]{0.9}$\mathbf{.650\pms{.002}}$ & \cellcolor[gray]{0.9}$\mathbf{.643\pms{.003}}$ & $.657\pms{.004}$ & \cellcolor[gray]{0.9} $\mathbf{.701\pms{.002}}$ & $.708\pms{.001}$ & $.681\pms{.002}$\\
\bottomrule
\multicolumn{9}{c}{}\\
\toprule
\parbox[t]{1.5mm}{\multirow{10}{*}{\rotatebox[origin=c]{90}{\textsc{Brier score} ($\downarrow$)}}}& Logistic Regression & $.131\pms{.000}$ & $.111\pms{.000}$ & $.051\pms{.000}$ & $.049\pms{.000}$ & $.101\pms{.000}$ & $.100\pms{.000}$ & $.087\pms{.000}$ & $.088\pms{.000}$\\
& Random Forest & $.133\pms{.000}$ & \cellcolor[gray]{0.9}$\mathbf{.110\pms{.000}}$ & $.055\pms{.000}$ & $.058\pms{.000}$ & $.104\pms{.000}$ & $.102\pms{.000}$ & \cellcolor[gray]{0.9}$\mathbf{.085\pms{.000}}$ & \cellcolor[gray]{0.9}$\mathbf{.086\pms{.000}}$\\
& MLP & $.133\pms{.001}$ & $.111\pms{.000}$ & $.050\pms{.000}$ & $.055\pms{.002}$ & $.103\pms{.000}$ & $.108\pms{.003}$ & $.087\pms{.000}$ & $.088\pms{.000}$\\
& Deep Ensemble & $.133\pms{.001}$ & \cellcolor[gray]{0.9}$\mathbf{.110\pms{.000}}$ & $.048\pms{.000}$ & $.052\pms{.001}$ & $.101\pms{.000}$ & $.100\pms{.000}$ & $.086\pms{.000}$ & $.086\pms{.000}$\\
\cmidrule(l){2-10}
& GIN & $.132\pms{.001}$ & $.112\pms{.001}$ & \multicolumn{2}{c|}{$.050\pms{.000}$} & \multicolumn{2}{c|}{$.103\pms{.000}$} & \multicolumn{2}{c}{$.090\pms{.001}$}\\
& GIN (attr masking) &\cellcolor[gray]{0.9} $\mathbf{.130\pms{.000}}$ & $.114\pms{.002}$ & \multicolumn{2}{c|}{$.049\pms{.000}$} & \multicolumn{2}{c|}{\cellcolor[gray]{0.9}$\mathbf{.100\pms{.000}}$} & \multicolumn{2}{c}{$.087\pms{.000}$}\\
& GIN (context pred) & $.134\pms{.000}$ & $.113\pms{.001}$ & \multicolumn{2}{c|}{$.050\pms{.000}$} & \multicolumn{2}{c|}{$.101\pms{.000}$} & \multicolumn{2}{c}{$.087\pms{.000}$}\\
& Grover & $.134\pms{.001}$ & $.111\pms{.001}$ & \multicolumn{2}{c|}{$.049\pms{.000}$} & \multicolumn{2}{c|}{$.101\pms{.000}$} & \multicolumn{2}{c}{$.088\pms{.000}$}\\
\cmidrule(l){2-10}
& \textbf{Q-SAVI} & \cellcolor[gray]{0.9}\cellcolor[gray]{0.9}$\mathbf{.130\pms{.000}}$ & $.112\pms{.003}$ & \cellcolor[gray]{0.9}$\mathbf{.047\pms{.000}}$ & \cellcolor[gray]{0.9}$\mathbf{.048\pms{.000}}$ & $.102\pms{.000}$ & \cellcolor[gray]{0.9}$\mathbf{.099\pms{.000}}$ & $.088\pms{.000}$ & $.090\pms{.000}$\\
\bottomrule
\end{tabular}
\end{adjustbox}
\vspace*{-10pt}
\end{table*}

\paragraph{Training and evaluation.}

To facilitate a fair comparison, we carried out an extensive hyperparameter search for every model, data split, and featurization. 
After an initial division of the data into training and test sets, the same data-splitting technique was applied again to derive a representative validation set.
The hyperparameter setting with the lowest negative log-likelihood on that validation set was then used to train ten independent models using different random seeds. Full implementation details and hyperparameter ranges are provided in~\Cref{appsec:training_details}.

Following model training and hyperparameter selection, the predictive accuracy and calibration of the estimated test-set label probabilities were characterized by the area under the \textsc{ROC} curve (\textsc{AUC-ROC}) and the \textsc{Brier Score}, as these enable the direct comparison of models across test sets with different label distributions (see~\Cref{tab:results}). Additionally, each algorithm's performance was characterized by the area under the precision-recall curve (\textsc{AUC-PRC}) and the adaptive calibration error (\textsc{ACE}; \citet{nixon2019measuring}), which closely mirror the \textsc{AUC-ROC} and \textsc{Brier Score} (see~\Cref{tab:results_auprc}). 

\paragraph{Results.}
The predictive accuracy and calibration metrics presented in~\Cref{tab:results,tab:results_auprc} (see~\Cref{app:sec_auprcs}) demonstrate that \textsc{q-savi} achieves significant performance gains in an out-of-distribution setting.
On the spectral and molecular weight splits---the evaluation settings with the strongest covariate and label shift---\textsc{q-savi} outperformed all other algorithms by a substantial and statistically significant margin in terms of predictive accuracy.
Similarly, its predictive uncertainty estimates were significantly better calibrated than all other algorithms on the molecular weight split and most other algorithms on the ECFP-based spectral split. 

On the substantially less data-shifted scaffold and random splits, relatively simple machine learning algorithms (e.g., random forests and deep ensembles) as well as more sophisticated self-supervised pre-training-based approaches consistently achieved the best predictive performance. 

In line with the empirical observations of \citet{ji2022drugood}, IRM, GroupDRO, DANN, and DeepCoral---domain adaptation and generalization techniques originally developed for images---were found to perform worse than most other techniques across most splits and featurizations (see~\Cref{tab:results_auprc}).

\subsection{Merck Molecular Activity Challenge}
\label{sec:merck-datasets}

As a complementary assessment of the practical utility of \textsc{q-savi}, we evaluated the method on the Merck Molecular Activity Challenge \citep{ma2015deep}. Consisting of 15 datasets from real-world production settings, it provides time-split training and test sets that represent the data shift encountered throughout a molecular optimization campaign \citep{sheridan2013time}. As the compound structures are only provided in the form of anonymized atom-pair descriptors in count and bit vector form, using a uniform subsample of a large chemical database as a context point distribution is not possible.

\setlength{\tabcolsep}{8.6pt}
\begin{table}[H]
    \vspace*{-20pt}
    \centering
    \caption{
    Covariate and label shift of time-split data from the Merck Molecular Activity Challenge.
    Covariate shift is quantified as the \mbox{(multi-)set} Jaccard kernel-based {\textsc{MMD}} statistic, while label shift is quantified as the two-sample Kolmogorov–Smirnov test statistic.}
    \label{tab:merck_cov_shift}
    \vspace{1pt}
    \small
    \begin{tabular}{lccc}
        \toprule
        \multirow{2}{*}{Dataset}& \multirow{2}{*}{Label Shift} & \multicolumn{2}{c}{Covariate Shift} \\
        & & Count Vector & Bit Vector\\
        \midrule
        \textsc{HIVPROT} & $0.579$ & $0.132$ & $0.162$ \\
        \textsc{DPP4} & $0.375$ & $0.112$ & $0.125$ \\
        \textsc{NK1} & $0.419$ & $0.071$ & $0.062$ \\
        \bottomrule
    \end{tabular}
\end{table}

\setlength{\tabcolsep}{12pt}
\begin{table*}[t]
\small
    \centering
    \caption{
    A summary of the test set performance of each model for each of the datasets from the Merck Molecular Activity Challenge, quantified by the mean squared error ($\downarrow$). All entries indicate the mean and standard error computed over $10$ independent training runs with different random seeds.
    The best models within a margin of statistical significance are highlighted in bold.}
    \label{tab:merck_results}
    \vspace{1pt}
    \begin{adjustbox}{max width=\textwidth}
    \begin{tabular}{l|cc|cc|cc}
    \toprule
    \small
    \multirow{2}{*}{Model} & \multicolumn{2}{c|}{HIVPROT} & \multicolumn{2}{c|}{\textsc{DPP4}} & \multicolumn{2}{c}{\textsc{NK1}}\\
    & count vector & bit vector & count vector & bit vector & count vector & bit vector \\
    \midrule
    $L_1$-Regression & $1.137\pms{.000}$ & $0.714\pms{.000}$ & $1.611\pms{.000}$ & $1.130\pms{.000}$ & $0.482\pms{.000}$ & $0.442\pms{.000}$ \\
    $L_2$-Regression & $0.999\pms{.000}$ & $0.723\pms{.000}$ & $1.495\pms{.000}$ & $1.143\pms{.000}$ & $0.498\pms{.000}$ & $0.436\pms{.000}$ \\
    Random Forest & $0.815\pms{.009}$ & $0.834\pms{.010}$ & $1.473\pms{.008}$ & $1.461\pms{.012}$ & $0.458\pms{.002}$ & $0.438\pms{.002}$ \\
    MLP & $0.768\pms{.014}$ & $2.118\pms{.015}$ & $1.393\pms{.024}$ & $1.094\pms{.029}$ & $0.443\pms{.007}$ & $0.399\pms{.006}$ \\
    \midrule
    \textbf{Q-SAVI} & \cellcolor[gray]{0.9}$\mathbf{0.682\pms{.019}}$ & \cellcolor[gray]{0.9}$\mathbf{0.664\pms{.028}}$ &\cellcolor[gray]{0.9} $\mathbf{1.332\pms{.017}}$ &\cellcolor[gray]{0.9} $\mathbf{1.028\pms{.027}}$ & \cellcolor[gray]{0.9}$\mathbf{0.436\pms{.007}}$ &\cellcolor[gray]{0.9} $\mathbf{0.387\pms{.012}}$ \\
    
    \bottomrule
    \end{tabular}
    \end{adjustbox}
\end{table*}

Instead, our evaluation focused on the three most covariate- and label-shifted datasets (see~\Cref{tab:merck_cov_shift}), repurposing the remaining data as an anonymized context point distribution.
All methods were evaluated following the protocol outlined in~\Cref{sec:experiments}, with full details presented in~\Cref{sec:app_merck}.
The performance metrics for our method and the baseline algorithms investigated in \citet{ma2015deep} are presented in~\Cref{tab:merck_results}, demonstrating that \textsc{q-savi} performs favorably across every setting and outperforms all other models on the strongly data-shifted \textsc{HIVPROT}, \textsc{DPP4}, and \textsc{NK1} datasets by a substantial and statistically significant margin.

\section{Summary and Conclusions}
\label{sec:discussion}

The objective of early-stage drug discovery is to identify lead compounds that exhibit sufficient evidence of modulating a given disease phenotype---as well as suitable safety profiles---to qualify them for further investigation in in-vivo studies.
Computational techniques that reliably predict the properties of novel molecules in unexplored regions of chemical space have the potential to substantially accelerate this time- and resource-intensive process.
Motivated by the practical importance of developing such methods, we derived \textsc{q-savi}, a probabilistic model that allows encoding explicit, problem-informed prior knowledge about the prediction domain into neural network training.

To construct a robust experimental setup and facilitate a practically meaningful evaluation of the proposed method, we carefully pre-processed a high-quality bioactivity dataset and explored different domain-specific statistics to quantify distribution shifts in this setting.
Using these statistics to highlight the limited extent to which commonly used random and scaffold splits are able to induce meaningful covariate and label shifts, we built on two alternative molecular weight- and spectral clustering-based approaches to construct challenging train-test splits.
Leveraging this extrapolative evaluation setup, we demonstrated that using \textsc{q-savi} to provide neural networks with relevant and contextualized information on drug-like chemical space significantly improves both the predictive accuracy and calibration of neural network models, outperforming a range of state-of-the-art self-supervised pre-training, ensembling, and domain adaptation techniques.

The main limitation of the proposed method compared to standard training regimes is its increased computational cost, due to the amortized cost of having to pre-process a suitable context point distribution and the direct cost of having to perform each forward pass over both a mini-batch and a sample of context points.
However, by keeping the size of each context set sample to be roughly comparable to the size of each mini-batch, we found this increase in computational cost to be manageable---especially in comparison to the computational cost of pre-training and fine-tuning related self-supervised methods or deep ensembles.

Promising avenues for future work include an investigation into how using \textsc{q-savi} to specify problem-informed modeling preferences may improve the performance of deep learning algorithms for drug discovery applications that heavily rely on out-of-distribution generalization.
For instance, the approach could be used to construct an acquisition function for an active learning loop to propose structural modifications that optimize the therapeutic properties of an existing lead compound~\citep{nicolaou2007molecular, gomez2018automatic}, as \textsc{q-savi} generates robust predictions and additionally enables researchers to explicitly specify desirable exit vectors.
It may also accelerate the discovery of novel compound classes that exhibit similar pharmacological properties to already explored molecules~\citep{bohm2004scaffold, hu2017recent}, enabling the optimization of certain pharmacokinetic properties or the circumvention of patent restrictions.
More broadly, we hope that this work encourages further research into the utility of probabilistic inference and domain-informed prior distributions over functions for drug discovery and beyond.
\vspace*{-6pt}
\section*{Acknowledgments}

We thank anonymous reviewers for useful feedback.
LK is funded by a Clarendon Scholarship.
TGJR is funded by a Qualcomm Innovation Fellowship.
We gratefully acknowledge the Oxford Advanced Research Computing service for providing computing resources and infrastructure.

\bibliography{references}
\bibliographystyle{apalike}

\clearpage

\begin{appendices}

\crefalias{section}{appsec}
\crefalias{subsection}{appsec}
\crefalias{subsubsection}{appsec}

\setcounter{equation}{0}
\renewcommand{\theequation}{\thesection.\arabic{equation}}

\onecolumn

\vspace*{-20pt}
{\hrule height 1mm}

\section*{\LARGE \centering Appendix
}
\label{sec:appendix}

\vspace{0.2in}
{\hrule height 0.3mm}
\vspace{14pt}

\section{Data Curation and Pre-Processing}
\label{appsec:dataset_processing}

 To generate an appropriate dataset of reliably labeled bioactivity measurements, we retrieved and reprocessed high-throughput screening data generated by \citet{antonova2018open} as part of a campaign to discover novel chemoprotective antimalarial drug candidates.
 
 The authors established a cell-based phenotypic screening pipeline to identify compounds that inhibit the development of luciferase-expressing liver-stage \textit{Plasmodium falciparum} parasites. After assaying a commercially-available chemical library of \num{538273} of drug-like small molecules in a single-point primary screen, they selected the \num{9963} most promising compounds for a series of confirmatory dose-response screens. Specifically, an 8-point dilution series was used to assess, in duplicate, the potency and efficacy of each compound in the original assay (\textit{Pbluc}). Additionally, the tendency of the assayed compounds to produce false positives and other experimental artifacts was investigated by performing a series of counter-screens that measure hepatic cytotoxicity (\textit{HepG2tox}) and interference with the luciferase-based luminescent readout (\textit{Ffluc}).
 The fact that all bioactivity measurements are (1) generated using biological duplicates and (2) associated with quantitative measures that reflect their likelihood to produce confounding experimental artifacts substantially improves the reliability of the resulting labels.

\begin{figure}[h]
\vspace*{-10pt}
    \centering
    \includegraphics[width=0.6\textwidth]{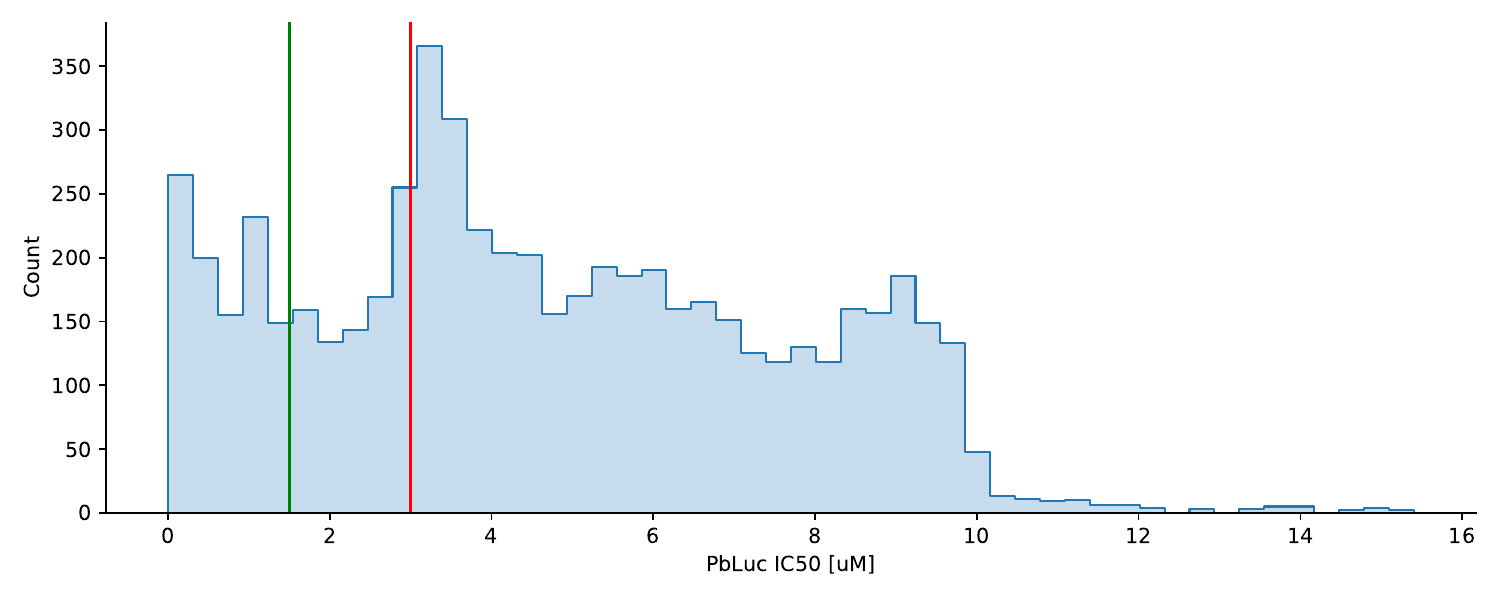}
    \caption{A histogram of the distribution of non-qualified \textit{Pbluc} $\textsc{IC}_{50}$ values. The green and red vertical lines indicate the thresholds (set to $\textsc{IC}_{50}=\num{1.5} \textrm{\textmu} \textrm{M}$ and $\textsc{IC}_{50}=\num{3} \textrm{\textmu} \textrm{M}$) that determine whether a compound is assigned to the active or inactive class.}
    \label{appfig:ic50}
\end{figure}

To facilitate the integration of bioactivity and counter-screen measurements and make the data more amenable to predictive modeling, the $\textsc{IC}_{50}$ values that quantify the concentration at which a molecule produces half of its maximum inhibitory effect were converted to binary labels.
Specifically, all compounds with an $\textsc{IC}_{50}\leq\num{1.5} \textrm{\textmu} \textrm{M}$ were denoted as active while all compounds with an $\textsc{IC}_{50}\geq\num{3} \textrm{\textmu} \textrm{M}$ were denoted as inactive, discarding \num{652} compounds with $\num{1.5} \textrm{\textmu} \textrm{M}\leq\textsc{IC}_{50}\leq\num{3} \textrm{\textmu} \textrm{M}$ and assigning qualified $\textsc{IC}_{50}$ values to the appropriate class (see~\Cref{appfig:ic50} for a diagram of the $\textsc{IC}_{50}$ distribution and the applied thresholds). 

In order to integrate information from the \textit{HepG2tox} and \textit{Ffluc} counter-screens and filter out problematic compounds that are likely false positives or risk confounding the evaluation in other ways, the thresholds outlined in \citet{antonova2018open} were applied. In particular, problematic compounds were discarded due to causing hepatotoxicity or assay interference if their respective $\textsc{IC}_{50}$ values met at least one of the criteria outlined in~\Cref{eq:filtering_criteria_a} and~\Cref{eq:filtering_criteria_b}
\begin{align}
    \label{eq:filtering_criteria_a}
    \textit{HepG2tox} \; \textsc{IC}_{50} < 2\cdot \textit{Pbluc}\;\textsc{IC}_{50}&\land\textit{HepG2tox} \; \textsc{IC}_{50} < c_{\text{max}} \\
    \label{eq:filtering_criteria_b}
    \textit{Ffluc} \; \textsc{IC}_{50} < 2\cdot \textit{Pbluc}\;\textsc{IC}_{50}&\land\textit{Ffluc} \; \textsc{IC}_{50} < c_{\text{max}},
\end{align}

where $c_\text{max}$ denotes the maximum concentration a compound was assayed at. These filtering criteria categorized \num{764} compounds as inhibiting hepatocyte viability and \num{446} compounds as interfering with the luminescence readout, including an overlap of \num{49}. Removing these compounds from the dataset results in a total of \num{8150} compounds, of which \num{7301} (90\%) are labeled as inactive and \num{849} (10\%) are labeled as active.

\newpage
\section{Additional Experimental Details}
\label{appsec:training_details}

This section provides additional implementational details of our experimental setup.
\Cref{app:sec_auprcs} presents the test set performance of each model under different data splitting and featurization techniques using the \textsc{AUC-PRC} and the \textsc{ACE} scores.
Additional experimental details are provided in~\Cref{app:impl_details}, describing the implementation and hyperparameter screening ranges for each model in our empirical evaluation, namely logistic regression (\Cref{app_sec:det_log_reg}), random forest classifiers (\Cref{app_sec:det_rf}), multi-layer perceptrons (\Cref{app_sec:det_mlp}), deep ensembles (\Cref{app_sec:det_deep_ens}), pre-trained graph neural networks (\Cref{app_sec:det_gnn}), GROVER (\Cref{app_sec:det_grover}), various domain adaptation and generalization techniques (\Cref{app_sec:det_domain_adapt}), and Q-SAVI (\Cref{app_sec:det_ours}). 
Detailed ablation plots that explore the influence of different hyperparameters on the performance of Q-SAVI are presented in~\Cref{app_sec:ablation_plots}.

All experiments and analyses were performed in Python \cite{van1995python}, using a range of general-purpose packages to aid model development and analysis \cite{harris2020array, Waskom2021, mckinney-proc-scipy-2010, virtanen2020scipy}.
All code that is necessary to reproduce the results presented in this work is available in the following repository: \href{https://github.com/leojklarner/Q-SAVI}{https://github.com/leojklarner/Q-SAVI}.

\subsection{Test Set Performances in \textsc{AUC-PRC} and \textsc{ACE}}
\label{app:sec_auprcs}

Following model training and hyperparameter selection, the predictive accuracy and calibration of the estimated test-set label probabilities were assessed using the area under the ROC curve (\textsc{AUC-ROC}) and the Brier score. These metrics allow for a direct comparison of models across test sets with different label distributions (see~\Cref{tab:results}). In addition, the performance of each algorithm was evaluated using the area under the precision-recall curve (\textsc{AUC-PRC}) and the adaptive calibration error (\textsc{ACE};~\citet{nixon2019measuring}). \textsc{AUC-PRC} and \textsc{ACE} are particularly well-suited for imbalanced datasets,  and provide a characterization of model performance that closely aligns with \textsc{AUC-ROC} and \textsc{Brier score} (see~\Cref{tab:results_auprc}). Note that their performance is not comparable across data splits, as it depends on the label distribution of a given test set---while the \textsc{AUC-ROC} of a no-skill classifier is $0.5$, its \textsc{AUC-PRC} is given by the positive label probability $p(y=1)$ of the test set.
\vspace{-0.5em}
\begin{table*}[h]
\centering
\caption{
    An overview of the test set performance of each model for each data splitting and featurization technique, quantified by the \textsc{AUC-PRC} ($\uparrow$) and the \textsc{ACE} ($\downarrow$) scores. All entries indicate the mean and standard error computed over $10$ independent training runs with different random seeds.
    The best models within a margin of statistical significance are highlighted in bold.
}
\small
\label{tab:results_auprc}
\begin{adjustbox}{max width=\textwidth}
\begin{tabular}{cl|cc|cc|cc|cc}
\toprule
 \multicolumn{2}{c|}{\multirow{2}{*}{Model \& Featurization}} & \multicolumn{2}{c|}{Spectral Split} & \multicolumn{2}{c|}{Weight Split} & \multicolumn{2}{c|}{Scaffold Split} & \multicolumn{2}{c}{Random Split}\\
 & & ECFP & rdkitFP &  ECFP & rdkitFP&  ECFP  & rdkitFP&  ECFP  & rdkitFP \\
\bottomrule
\multicolumn{9}{c}{}\\
\toprule
\parbox[t]{1.5mm}{\multirow{14}{*}{\rotatebox[origin=c]{90}{\textsc{AUC-PRC} ($\uparrow$)}}} 
& Logistic Regression & $.211\pms{.000}$ & $.140\pms{.000}$ & $.106\pms{.000}$ &  \cellcolor[gray]{0.9}$\mathbf{.112\pms{.000}}$ & $.211\pms{.000}$ & $.211\pms{.000}$ & $.248\pms{.000}$ & $.225\pms{.000}$\\
& Random Forest & $.207\pms{.005}$ & $.141\pms{.002}$ & $.090\pms{.002}$ & $.089\pms{.003}$ & $.165\pms{.002}$ & $.200\pms{.002}$ & \cellcolor[gray]{0.9} $\mathbf{.292\pms{.002}}$ & \cellcolor[gray]{0.9} $\mathbf{.294\pms{.002}}$\\
& MLP & $.208\pms{.003}$ & $.144\pms{.001}$ & $.090\pms{.003}$ & $.089\pms{.004}$ & $.180\pms{.005}$ & $.184\pms{.007}$ & $.270\pms{.004}$ & $.233\pms{.006}$\\
& Deep Ensemble & $.217\pms{.003}$ & $.144\pms{.001}$ & $.114\pms{.001}$ & $.102\pms{.003}$ & $.209\pms{.002}$ & $.209\pms{.002}$ & $.288\pms{.002}$ & $.266\pms{.004}$\\
\cmidrule(l){2-10}
& GIN & $.183\pms{.006}$ & $.149\pms{.005}$ & \multicolumn{2}{c|}{$.082\pms{.004}$} & \multicolumn{2}{c|}{$.202\pms{.005}$} & \multicolumn{2}{c}{$.218\pms{.005}$}\\
& GIN (attr masking) & $.192\pms{.003}$ & $.152\pms{.005}$ & \multicolumn{2}{c|}{$.108\pms{.003}$} & \multicolumn{2}{c|}{\cellcolor[gray]{0.9} $\mathbf{.245\pms{.004}}$} & \multicolumn{2}{c}{$.251\pms{.003}$}\\
& GIN (context pred) & $.188\pms{.004}$ & $.152\pms{.004}$ & \multicolumn{2}{c|}{$.098\pms{.002}$} & \multicolumn{2}{c|}{$.206\pms{.005}$} & \multicolumn{2}{c}{$.254\pms{.005}$}\\
& Grover & $.199\pms{.001}$ & $.139\pms{.002}$ & \multicolumn{2}{c|}{$.106\pms{.001}$} & \multicolumn{2}{c|}{$.204\pms{.004}$} & \multicolumn{2}{c}{$.227\pms{.003}$}\\
\cmidrule(l){2-10}
& IRM & $.154\pms{.004}$ &$.145\pms{.004}$&\multicolumn{2}{c|}{$.086\pms{.004}$}&\multicolumn{2}{c|}{$.178\pms{.005}$}&\multicolumn{2}{c}{$.176\pms{.004}$}\\
& GroupDRO &$.166\pms{.003}$&$.159\pms{.002}$&\multicolumn{2}{c|}{$.102\pms{.003}$}&\multicolumn{2}{c|}{$.202\pms{.003}$}&\multicolumn{2}{c}{$.172\pms{.005}$}\\
& DANN &$.156\pms{.003}$&$.155\pms{.005}$&\multicolumn{2}{c|}{$.095\pms{.005}$}&\multicolumn{2}{c|}{$.184\pms{.004}$}&\multicolumn{2}{c}{$.202\pms{.003}$}\\
& DeepCoral &$.154\pms{.004}$&$.151\pms{.004}$&\multicolumn{2}{c|}{$.091\pms{.003}$}&\multicolumn{2}{c|}{$.194\pms{.003}$}&\multicolumn{2}{c}{$.212\pms{.003}$}\\
\cmidrule(l){2-10}
& Q-SAVI & \cellcolor[gray]{0.9}$\mathbf{.221\pms{.003}}$ &  \cellcolor[gray]{0.9}$\mathbf{.165\pms{.004}}$ &  \cellcolor[gray]{0.9}$\mathbf{.121\pms{.002}}$ &  \cellcolor[gray]{0.9}$\mathbf{.111\pms{.003}}$ & $.197\pms{.003}$ & $.216\pms{.003}$ & $.239\pms{.002}$ & $.208\pms{.004}$\\
\bottomrule
\multicolumn{9}{c}{}\\
\toprule
\parbox[t]{1.5mm}{\multirow{14}{*}{\rotatebox[origin=c]{90}{\textsc{ACE} ($\downarrow$)}}}
& Logistic Regression & $.061\pms{.000}$ & $.055\pms{.000}$ & $.041\pms{.000}$ & $.034\pms{.000}$ & $.026\pms{.000}$ &  \cellcolor[gray]{0.9}$\mathbf{.025\pms{.000}}$ & $.018\pms{.000}$ & $.024\pms{.000}$\\
& Random Forest & $.078\pms{.001}$ &  \cellcolor[gray]{0.9}$\mathbf{.033\pms{.001}}$ & $.074\pms{.001}$ & $.087\pms{.001}$ & $.029\pms{.001}$ &  \cellcolor[gray]{0.9}$\mathbf{.025\pms{.001}}$ &  \cellcolor[gray]{0.9}$\mathbf{.016\pms{.001}}$ & $.035\pms{.001}$\\
& MLP & $.079\pms{.003}$ & $.052\pms{.003}$ & $.035\pms{.003}$ & $.055\pms{.007}$ & $.029\pms{.002}$ & $.044\pms{.011}$ & $.029\pms{.001}$ & $.026\pms{.002}$\\
& Deep Ensemble & $.078\pms{.004}$ & $.050\pms{.001}$ & $.025\pms{.001}$ & $.053\pms{.005}$ &  \cellcolor[gray]{0.9}$\mathbf{.022\pms{.001}}$ &  \cellcolor[gray]{0.9}$\mathbf{.025\pms{.001}}$ & $.023\pms{.001}$ & $.019\pms{.001}$\\
\cmidrule(l){2-10}
& GIN & $.064\pms{.004}$ & $.047\pms{.007}$ & \multicolumn{2}{c|}{$.036\pms{.003}$} & \multicolumn{2}{c|}{$.033\pms{.003}$} & \multicolumn{2}{c}{$.026\pms{.003}$}\\
& GIN (attr masking) &  \cellcolor[gray]{0.9}$\mathbf{.053\pms{.002}}$ & $.057\pms{.009}$ & \multicolumn{2}{c|}{$.038\pms{.002}$} & \multicolumn{2}{c|}{$.030\pms{.001}$} & \multicolumn{2}{c}{$.020\pms{.001}$}\\
& GIN (context pred) & $.078\pms{.002}$ & $.051\pms{.005}$ & \multicolumn{2}{c|}{$.034\pms{.003}$} & \multicolumn{2}{c|}{$.028\pms{.002}$} & \multicolumn{2}{c}{\cellcolor[gray]{0.9}$\mathbf{.015\pms{.001}}$}\\
& Grover & $.074\pms{.004}$ & \cellcolor[gray]{0.9} $\mathbf{.035\pms{.002}}$ & \multicolumn{2}{c|}{$.036\pms{.002}$} & \multicolumn{2}{c|}{$.038\pms{.002}$} & \multicolumn{2}{c}{$.020\pms{.001}$}\\
\cmidrule(l){2-10}
& IRM & $.071\pms{.003}$ &$.067\pms{.002}$&\multicolumn{2}{c|}{$.044\pms{.002}$}&\multicolumn{2}{c|}{$.035\pms{.001}$}&\multicolumn{2}{c}{$.024\pms{.002}$}\\
& GroupDRO &$.060\pms{.003}$&\cellcolor[gray]{0.9}$\mathbf{.035\pms{.003}}$&\multicolumn{2}{c|}{$.039\pms{.002}$}&\multicolumn{2}{c|}{$.036\pms{.002}$}&\multicolumn{2}{c}{$.026\pms{.001}$}\\
& DANN &$.057\pms{.002}$&$.046\pms{.003}$&\multicolumn{2}{c|}{$.035\pms{.003}$}&\multicolumn{2}{c|}{$.028\pms{.001}$}&\multicolumn{2}{c}{$.030\pms{.002}$}\\
& DeepCoral &$.097\pms{.006}$&\cellcolor[gray]{0.9}$\mathbf{.035\pms{.002}}$&\multicolumn{2}{c|}{$.041\pms{.004}$}&\multicolumn{2}{c|}{$.036\pms{.002}$}&\multicolumn{2}{c}{$.026\pms{.002}$}\\
\cmidrule(l){2-10}
& Q-SAVI &  \cellcolor[gray]{0.9}$\mathbf{.052\pms{.001}}$ & $.043\pms{.013}$ &  \cellcolor[gray]{0.9}$\mathbf{.015\pms{.001}}$ &  \cellcolor[gray]{0.9}$\mathbf{.016\pms{.001}}$ & $.036\pms{0
.002}$ &  \cellcolor[gray]{0.9}$\mathbf{.025\pms{.002}}$ & $.021\pms{.001}$ & $.024\pms{.002}$\\
\bottomrule
\end{tabular}
\end{adjustbox}
\end{table*}

\clearpage

\subsection{Model Implementations and Hyperparameter Ranges}
\label{app:impl_details}

To ensure a fair and meaningful comparison of the evaluated machine learning models, the hyperparameters of each algorithm were independently optimized for every data split and featurization technique. The following sections provide comprehensive details about the implementation and hyperparameter ranges used for each model in our empirical evaluation.

\begin{itemize}
\setlength{\itemsep}{-2pt}
\item Logistic Regression (Section~\ref{app_sec:det_log_reg})
\item Random Forest Classifiers (Section~\ref{app_sec:det_rf})
\item Multi-layer Perceptrons (Section~\ref{app_sec:det_mlp})
\item Deep Ensembles (Section~\ref{app_sec:det_deep_ens})
\item Pre-trained Graph Neural Networks (Section~\ref{app_sec:det_gnn})
\item GROVER (Section~\ref{app_sec:det_grover})
\item Domain Adaptation and Generalization Techniques (Section~\ref{app_sec:det_domain_adapt})
\item Our Probabilistic Regularization Scheme (Section~\ref{app_sec:det_ours})
\end{itemize}

\subsubsection{Logistic Regression}
\label{app_sec:det_log_reg}

The \textbf{logistic regression models} were trained with the scikit-learn library \cite{scikit-learn} using the \textsc{liblinear} solver \cite{fan2008liblinear} with a maximum of 1000 iterations and a stopping tolerance of \num{1e-4}. They were independently fit for all hyperparameter combinations specified in\Cref{tab:hyper_logreg}, using the combination with the best unweighted validation set log-likelihood to choose the best hyperparameter setting to evaluate on the held-out test set.

\setlength{\tabcolsep}{20.0pt}
\begin{table}[H]
\centering
\caption{Hyperparameters for Logistic Regression}
\label{tab:hyper_logreg}
\vspace{0.3em}
\begin{tabular}{lll}
\toprule
\textbf{Model} & \textbf{Hyperparameter} & \textbf{Search Space} \\
\midrule
Linear Regression & regularization type & $\ell1$, $\ell2$ \\
 & regularization strength & \num{1.0e-4}, \num{2.6e-04} \ldots, \num{3.8e+03}, \num{1.0e+04} \\
 & class weight & none, balanced \\
\bottomrule
\end{tabular}
\end{table}

\subsubsection{Random Forest Classifiers}
\label{app_sec:det_rf}
 
The \textbf{random forest models} were trained with the scikit-learn library \cite{scikit-learn} using 100 decision trees and the \textsc{gini} splitting criterion.
They were independently fit for all hyperparameter combinations specified in~\Cref{tab:rfr_hypers}, using the combination with the best unweighted validation set log-likelihood to choose the best hyperparameter setting to evaluate on the held-out test set.

\setlength{\tabcolsep}{32.0pt}
\begin{table}[H]
\centering
\caption{Hyperparameters for Random Forest Classifiers}
\label{tab:rfr_hypers}
\vspace{0.3em}
\begin{tabular}{lll}
\toprule
\textbf{Model} & \textbf{Hyperparameter} & \textbf{Search Space} \\
\midrule
Random Forest & maximum depth & 5, 15, 26, 36, 47, 57, 68, 78, 89, 100 \\
 & min. samples per split & 5, 15, 50, 100 \\
 & min. samples per leaf & 1, 5, 10, 30, 100 \\
 & class weight & none, balanced \\
\bottomrule
\end{tabular}
\end{table}

\clearpage

\subsubsection{Multi-Layer Perceptrons}
\label{app_sec:det_mlp}

The \textbf{multi-layer perceptrons} were implemented with the PyTorch library \cite{paszke2019pytorch}, using rectified linear units \cite{nair2010rectified} as activation functions. Their weights were initialized using a Normal distribution $\mathcal{N}(0,1)$ truncated at $\pm2\sigma$, with biases initialized at zero. These parameters were optimized on the training set using the \textsc{AdamW} stochastic gradient descent optimizer \cite{loshchilov2017decoupled} with a batch size of 128 and the cross-entropy loss for a maximum of 500 epochs, using early stopping to terminate training if the unweighted log-likelihood on the validation set did not decrease for more than 10 epochs, reverting to the checkpoint with best validation set log-likelihood for evaluating their performance for hyperparameter optimization and the subsequent on the held-out test set. Batch normalization and dropout were applied after the ReLU non-lineary. The full hyperparameter search space is presented in~\Cref{table:mlp_hypers}.

\setlength{\tabcolsep}{28.0pt}
\begin{table}[H]
\centering
\caption{Hyperparameters for Multi-layer Perceptrons}
\label{table:mlp_hypers}
\vspace{0.3em}
\begin{tabular}{lll}
\toprule
\textbf{Model} & \textbf{Hyperparameter} & \textbf{Search Space} \\
\midrule
Multi-layer Perceptron & learning rate & \num{1e-4}, \num{1e-3} \\
 & weight decay & \num{1e-3}, \num{1e-2}, \num{1e-1} \\
 & number of layers & 2, 4, 6 \\
 & embedding dimension & 32, 64 \\
 & batch normalization (BN) & yes, no \\
 & BN running statistics & yes, no \\
 & dropout & 0.0, 0.2, 0.5 \\
 & class weight & none, balanced \\
\bottomrule
\end{tabular}
\end{table}

\subsubsection{Deep Ensembles}
\label{app_sec:det_deep_ens}

The \textbf{deep ensembles} were trained using an identical setup to the multi-layer perceptrons, with the distinction that $M=5$ independent networks were trained with different random seeds and evaluated with respect to their average log-likelihood on the validation set. Similarly, at inference time the class probabilities were averaged across ensembles. The full hyperparameter search space is presented in~\Cref{tab:deep_ensembles_hypers} and is identical to~\Cref{table:mlp_hypers}.

\setlength{\tabcolsep}{33.0pt}
\begin{table}[H]
\centering
\caption{Hyperparameters for Deep Ensembles}
\label{tab:deep_ensembles_hypers}
\vspace{0.3em}
\begin{tabular}{lll}
\toprule
\textbf{Model} & \textbf{Hyperparameter} & \textbf{Search Space} \\
\midrule
Deep Ensemble & learning rate & \num{1e-4}, \num{1e-3} \\
 & weight decay & \num{1e-3}, \num{1e-2}, \num{1e-1} \\
 & number of layers & 2, 4, 6 \\
 & embedding dimension & 32, 64 \\
 & batch normalization (BN) & yes, no \\
 & BN running statistics & yes, no \\
 & dropout & 0.0, 0.2, 0.5 \\
 & class weight & none, balanced \\
\bottomrule
\end{tabular}
\end{table}

\clearpage

\subsubsection{Pre-trained Graph Neural Networks}
\label{app_sec:det_gnn}

The graph featurization pipeline, architectures,  and pre-trained initializations of the graph isomorphism networks presented in~\citet{hu2019strategies} were retrieved from the paper's official GitHub repository and fine-tuned on the training set using the \textsc{AdamW} optimizer \cite{loshchilov2017decoupled} with a batch size of 128 and the cross-entropy loss for a maximum of 500 epochs, using early stopping to terminate training if the unweighted log-likelihood on the validation set did not decrease for more than 10 epochs and reverting to the checkpoint with best validation set log-likelihood for evaluating their performance for hyperparameter optimization and the subsequent on the held-out test set. The full hyperparameter search space is presented in~\Cref{tab:pre_trained_hypers}. The pre-trained initializations were provided for networks with $5$ layers of $300$ hidden units, set up using batch normalization with running statistics.

\setlength{\tabcolsep}{22.5pt}
\begin{table}[H]
\centering
\caption{Hyperparameters for Pre-trained GINs}
\label{tab:pre_trained_hypers}
\vspace{0.3em}
\begin{tabular}{lll}
\toprule
\textbf{Model} & \textbf{Hyperparameter} & \textbf{Search Space} \\
\midrule
Pre-trained GINs & learning rate & \num{1e-4}, \num{3e-3}, \num{1e-3}, \num{3e-3}, \num{1e-2} \\
 & weight decay & \num{1e-3}, \num{1e-2}, \num{1e-1} \\
 & dropout & 0.0, 0.2, 0.5 \\
 & class weight & none, balanced \\
\bottomrule
\end{tabular}
\end{table}

\subsubsection{GROVER}
\label{app_sec:det_grover}

All code, models, and initializations required to fine-tune the pre-trained graph transformers presented in~\citet{rong2020self} was retrieved from the paper's official GitHub repository and fine-tuned on the training set with a batch size of 128 for a maximum of 500 epochs, using early stopping to terminate training if the unweighted log-likelihood on the validation set did not decrease for more than 10 epochs and reverting to the checkpoint with best validation set log-likelihood for evaluating their performance for hyperparameter optimization and the held-out test set. The hyperparameters specifying the number of layers and their embedding dimension indicate the size of the MLP fit on top of the pre-trained molecular representations produced by the \textsc{GROVER} base model and were chosen to be identical to the other MLP-based deep learning algorithms. The full hyperparameter search space is presented in~\Cref{tab:grover_hypers}.

\setlength{\tabcolsep}{38.5pt}
\begin{table}[H]
\centering
\caption{Hyperparameters for \textsc{GROVER}}
\label{tab:grover_hypers}
\vspace{0.3em}
\begin{tabular}{lll}
\toprule
\textbf{Model} & \textbf{Hyperparameter} & \textbf{Search Space} \\
\midrule
\textsc{GROVER} & learning rate & \num{1e-4}, \num{1e-3} \\
 & weight decay & \num{1e-3}, \num{1e-2}, \num{1e-1} \\
 & dropout & 0.0, 0.2, 0.5 \\
 & number of layers & 2, 4, 6 \\
 & embedding dimension & 32, 64 \\
\bottomrule
\end{tabular}
\end{table}

\clearpage

\subsubsection{Domain Adaptation Techniques}
\label{app_sec:det_domain_adapt}

All code and featurization utilities required to run the evaluated domain adaptation and generalization techniques, namely invariant risk minimization (\textbf{IRM}), group-distributionally robust training (\textbf{GroupDRO}), domain-adversarial networks (\textbf{DANN}) and deep correlation alignment (\textbf{DeepCoral}), were adapted from \citet{ji2022drugood} and provided with data split-specific domain indicators. For this, the training set was additionally split into three domains, either using spectral clustering, molecular weight thresholds, a grouped scaffold split, or random partitions. 
All models used the default architecture choice in~\citet{ji2022drugood}---a graph isomorphism network with 4 layers and 128 hidden units---and trained according to the respective optimization procedures. The full hyperparameter range is presented in~\Cref{tab:domain_adapt_hypers}.

\setlength{\tabcolsep}{23.5pt}
\begin{table}[H]
\centering
\caption{Hyperparameters for Domain Adaptation Techniques}
\label{tab:domain_adapt_hypers}
\vspace{0.3em}
\begin{tabular}{lll}
\toprule
\textbf{Model} & \textbf{Hyperparameter} & \textbf{Search Space} \\
\midrule
IRM/GroupDRO/DANN/DeepCoral & learning rate & \num{1e-4}, \num{1e-3} \\
 & weight decay & \num{1e-3}, \num{1e-2}, \num{1e-1} \\
 & dropout & 0.0, 0.2, 0.5 \\
\bottomrule
\end{tabular}
\end{table}

\subsubsection{Q-SAVI}
\label{app_sec:det_ours}

The models based on our probabilistic regularization scheme were trained using the implementation of the local linearization scheme presented in~\citet{Rudner2022fsvi, Rudner2022sfsvi} provided by the authors and using the exact same architecture, initialization, and optimization procedures as for the multi-layer perceptrons and deep ensembles---differing only in the objective function. 
Specifically, at each gradient step iteration, a sample of $M$ molecules (where $M$ is a hyperparameter) was drawn from a uniform distribution over the ZINC database~\citep{irwin2020zinc20}, providing a set of context points on which to evaluate the objective in~\Cref{eq:objective}, using the Bernoulli likelihood to specify $\log p(\by_{\calD} \vbar f(\bx_{\calD} ; \btheta))$.
To construct a prior distribution over parametric function mappings $p_{f(\{ \bX, \bX_{\calC} \} ; \bTheta)}$ that maximizes predictive uncertainty away from the training data, it was defined as a distribution over functions with a logit-space mean vector of approximately zero and minimal structure in the off-diagonal entries of its covariance matrix. We refer to our code repository for further implementational details. The full hyperparameter search space is presented in~\Cref{tab:ours_hypers}.

\setlength{\tabcolsep}{31.0pt}
\begin{table}[H]
\centering
\caption{Hyperparameters for Our Model}
\label{tab:ours_hypers}
\vspace{0.3em}
\begin{tabular}{lll}
\toprule
\textbf{Model} & \textbf{Hyperparameter} & \textbf{Search Space} \\
\midrule
Q-SAVI & learning rate & \num{1e-4}, \num{1e-3} \\
 & number of layers & 2, 4, 6 \\
 & embedding dimension & 32, 64 \\
 & prior variance & \num{1e-2}, \num{1e-1}, \num{1e0}, \num{1e1}, \num{1e2} \\
 & context points per sample & 16, 128 \\
\bottomrule
\end{tabular}
\end{table}

\clearpage

\subsection{Ablation Studies}
\label{app_sec:ablation_plots}

\begin{figure}[H]
    \centering
    \hfill
    \begin{subfigure}[b]{0.44\textwidth}
        \centering
        \includegraphics[width=\textwidth]{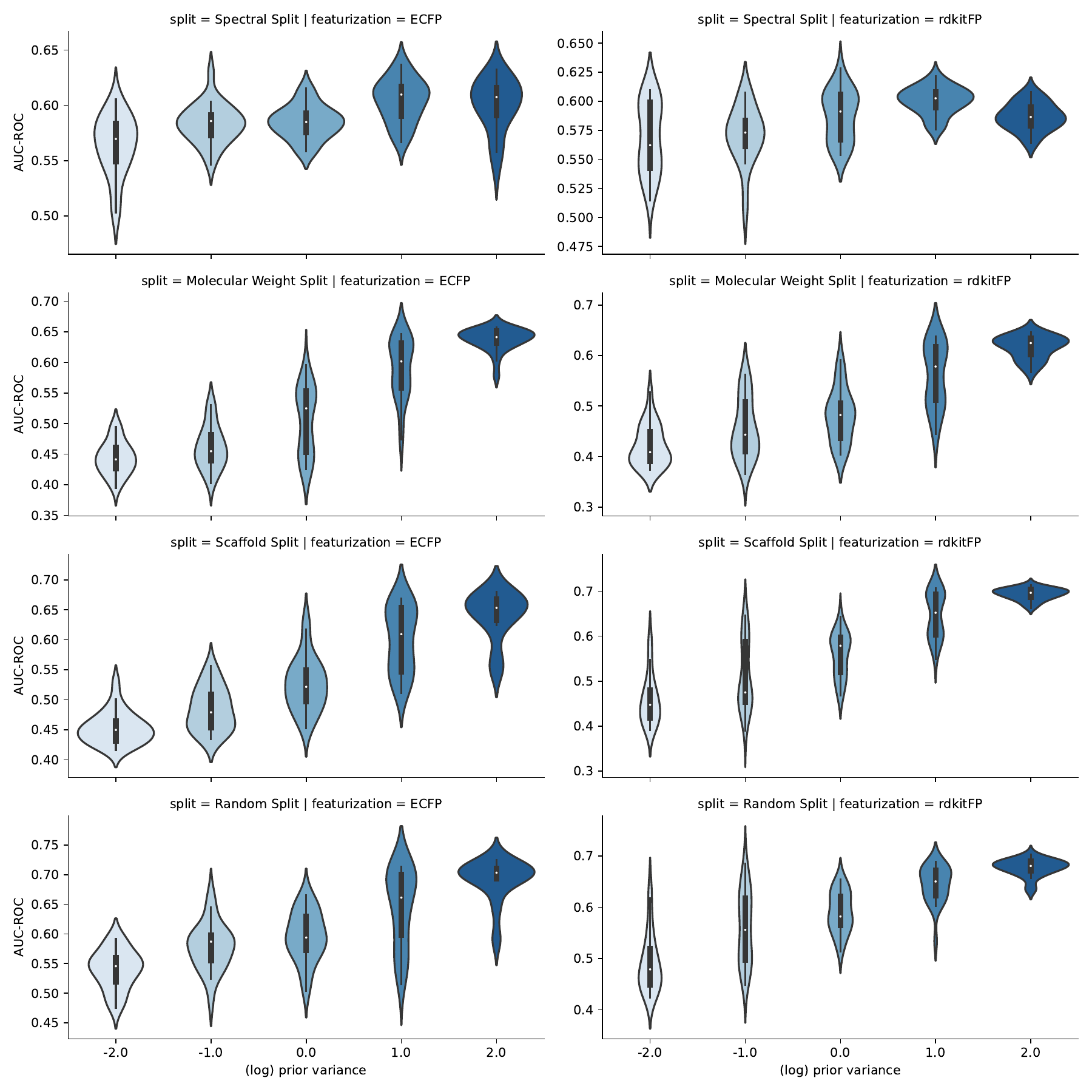} %
        \caption{Ablation plot showing the effect of the prior covariance on the test set \textsc{AUC-ROC}.} %
        \label{fig:prior_cov_auc_roc} %
    \end{subfigure}
    \hfill
    \begin{subfigure}[b]{0.44\textwidth}
        \centering
        \includegraphics[width=\textwidth]{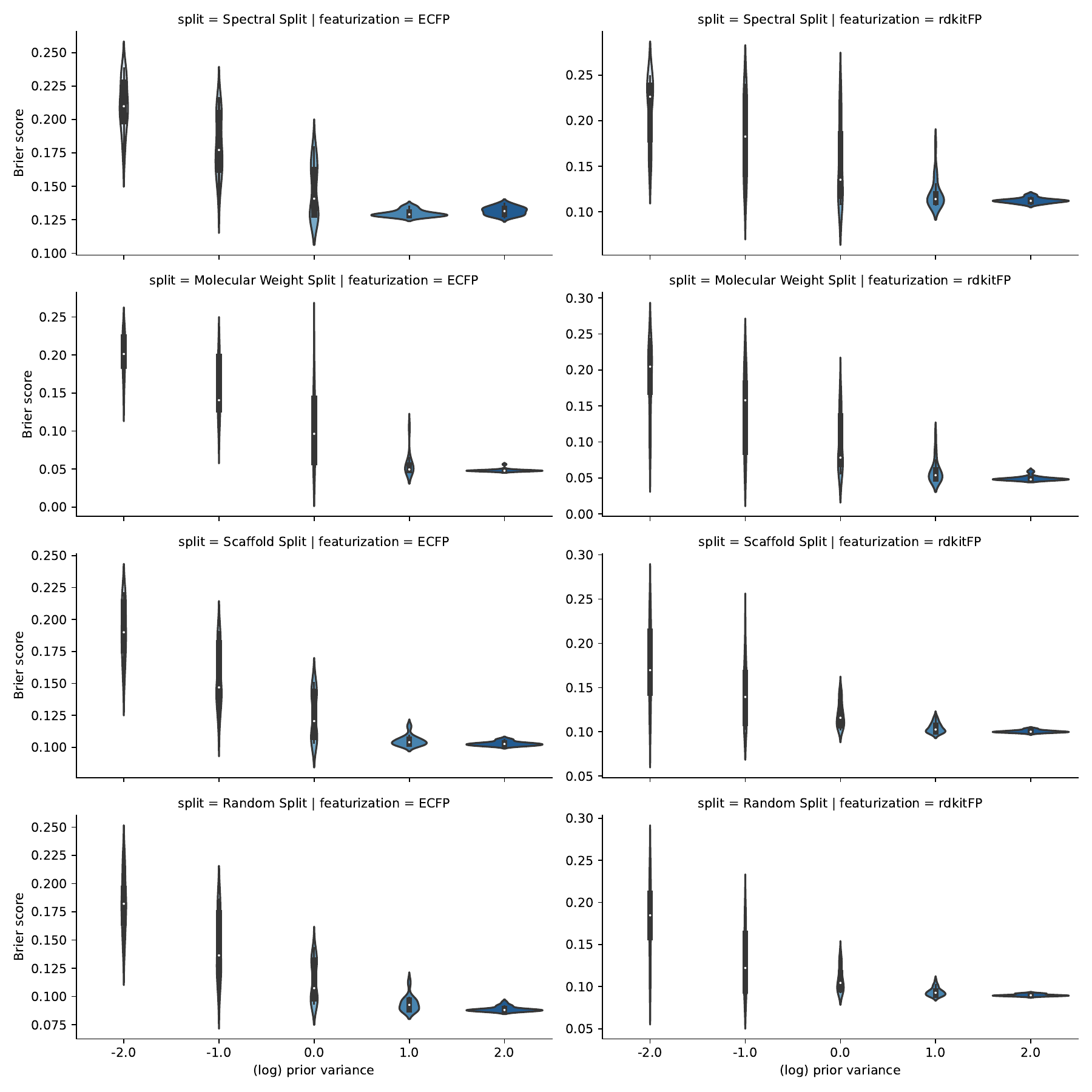} %
        \caption{Ablation plot showing the effect of the prior covariance on the test set \textsc{Brier score}.} %
        \label{fig:prior_cov_brier_score} %
    \end{subfigure}
    \hfill
    \caption{Effect of (log) prior variance on the test set performance metrics.} %
    \label{fig:prior_cov_effect} %
\end{figure}

\begin{figure}[H]
    \centering
    \hfill
    \begin{subfigure}[b]{0.43\textwidth}
        \centering
        \includegraphics[width=\textwidth]{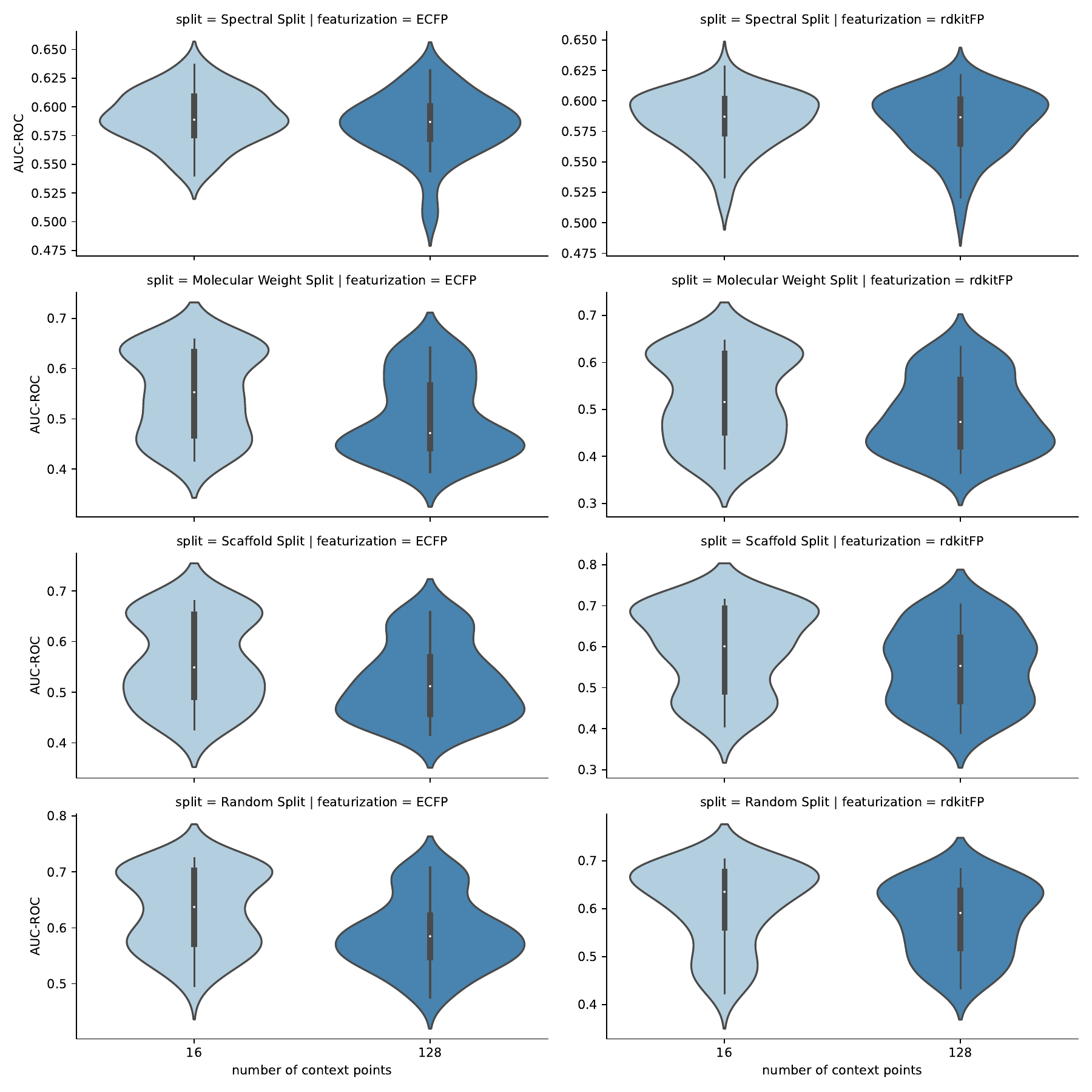} %
        \caption{Ablation plot showing the effect of the number of context points on the test set \textsc{AUC-ROC}.} %
        \label{fig:n_inducing_inputs_auc_roc} %
    \end{subfigure}
    \hfill
    \begin{subfigure}[b]{0.43\textwidth}
        \centering
        \includegraphics[width=\textwidth]{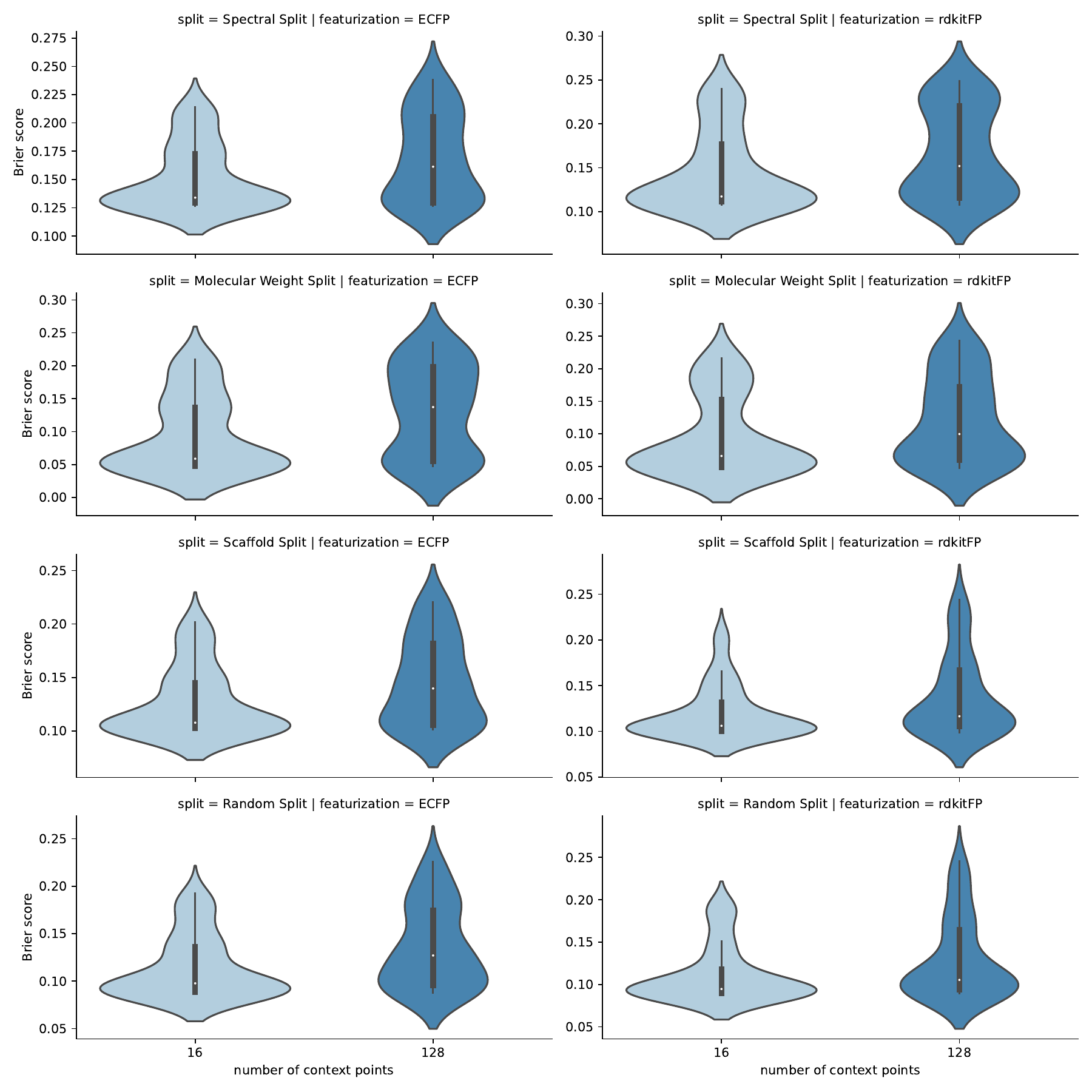} %
        \caption{Ablation plot showing the effect of the number of context points on the test set \textsc{Brier score}.} %
        \label{fig:n_inducing_inputs_brier_score} %
    \end{subfigure}
    \hfill
    \caption{Effect of the number of context points on the test set performance metrics.} %
    \label{fig:n_inducing_inputs_effect} %
\end{figure}

To understand the impact of different hyperparameters on the performance of our proposed probabilistic regularization scheme, we conducted a series of ablation experiments. In these experiments, we systematically varied the hyperparameters relevant to evaluating the objective in~\Cref{eq:objective}---namely the prior variance and the number of sampled context points---while keeping others fixed, and measured their effects on the test set \textsc{AUC-ROC} and \textsc{Brier Score}. The resulting ablation plots are presented in~\Cref{fig:prior_cov_effect,fig:n_inducing_inputs_effect} and show that larger prior covariances are strongly correlated with more robust test-set performances across splits---while the effect of larger context point samples is less pronounced.

\clearpage

\section{Additional Experimental Details for the Merck Molecular Activity Challenge Data}
\label{sec:app_merck}

\begin{figure}[H]
    \centering
    \includegraphics[width=0.8\textwidth]{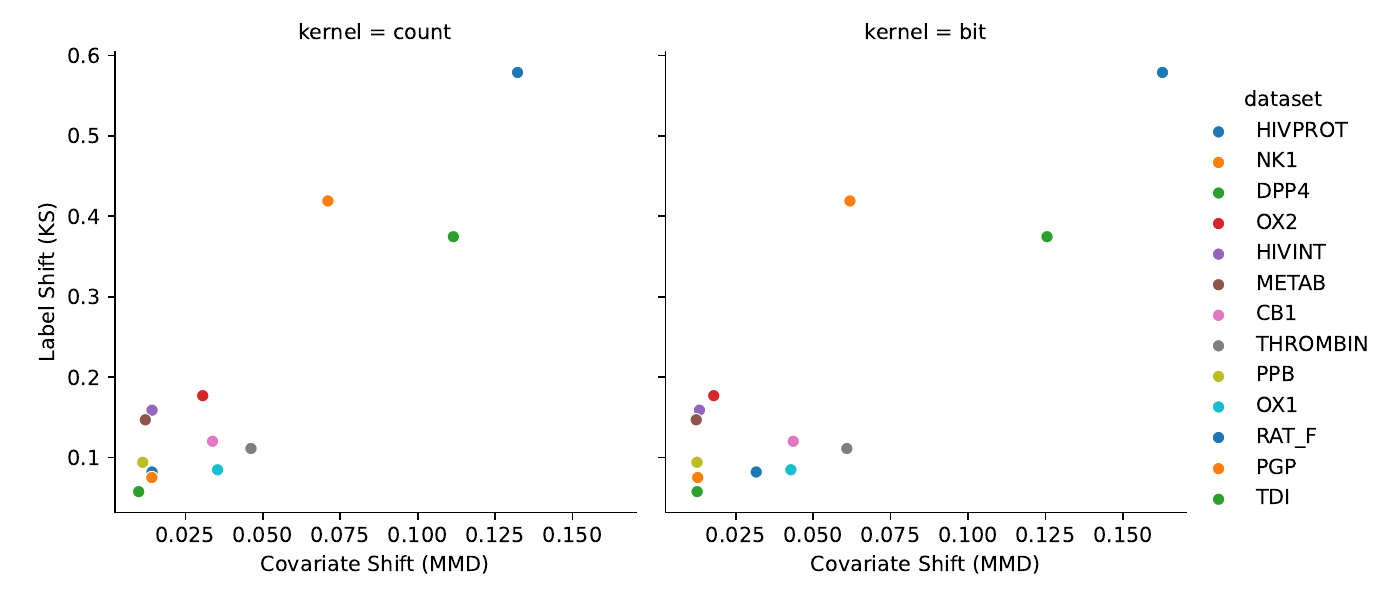} %
    \caption{Scatterplot illustrating the covariate and label shifts in the Merck Molecular Activity Challenge datasets} %
    \label{fig:covariate_label_shift_scatterplot} %
\end{figure}

In order to assess the practical utility of our method in real-world production settings, an evaluation on the Merck Molecular Activity Challenge datasets \cite{ma2015deep} was conducted. This data consists of 15 datasets from real-world production environments with time-split training and test sets. As the compound structures are only provided in the form of anonymized atom-pair descriptors, it is not possible to use a uniform subsample of a large chemical database as a context point distribution.
Instead, the evaluation focused on the three most covariate- and label-shifted datasets, see~\Cref{fig:covariate_label_shift_scatterplot}, using a uniform distribution over molecules from the remaining datasets as a context point distribution. To select these datasets, the multiset version of the standard Jaccard/Tanimoto index
$$
k_{\text{jac-multiset}}(\mathbf{x}, \mathbf{y}) = \frac{\sum_{i} \min(\mathbf{x}_i, \mathbf{y}_i)}{\sum_{i} \max(\mathbf{x}_i, \mathbf{y}_i)}
$$
was used to evaluate the \textsc{MMD} statistic between two sets of count vectors and quantify covariate shift. Label shift between the regression targets of every training and test set was quantified through the two-sample Kolmogorov-Smirnov test statistic.

\vspace*{10pt}

\begin{wrapfigure}{r}{0.5\linewidth}
    \centering
    \vspace*{-20pt}
    \includegraphics[width=\linewidth]{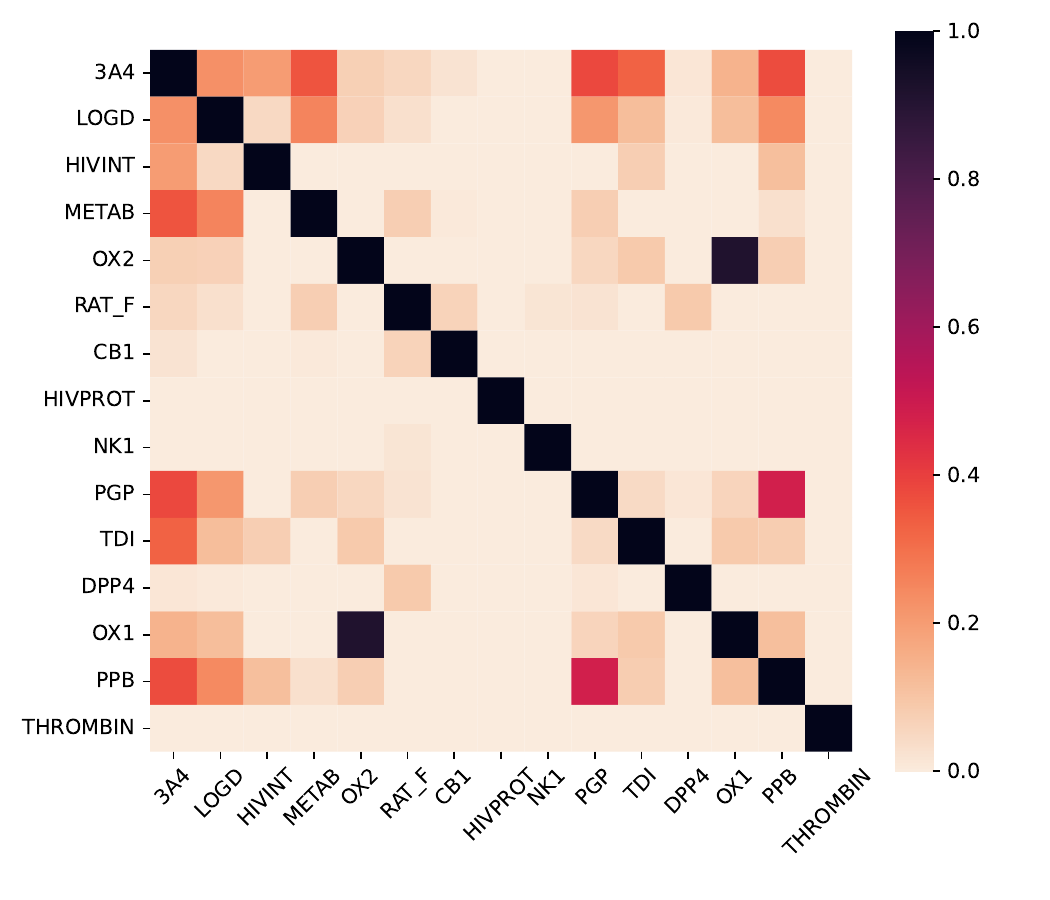} %
    \vspace*{-30pt}
    \caption{Heatmap illustrating the pairwise overlap between different datasets from the Merck Molecular Activity Challenge, defined as the proportion of molecules from the smaller dataset that are found in the larger dataset, i.e., ${|\mathbf{X}_1\cup\mathbf{X}_2|}/{\min(|\mathbf{X}_1|,|\mathbf{X}_2|)}$} %
    \label{fig:merck_dataset_overlap} %
\end{wrapfigure}

As shown in~\Cref{fig:merck_dataset_overlap}, the direct overlap between the \textsc{HIVPROT}, \textsc{NK1}, and \textsc{DPP4} datasets with the remaining data is minimal, warranting its use as a general and diverse context point distribution.
Using this evaluation setup, 10\% of the training sets was randomly split off as a validation set for hyperparameter optimization and, where applicable, early stopping. Model-specific details are outlined below, including implementational details and hyperparameter ranges for regularized linear regressions (\Cref{app_sec:det_lin_reg_regression}), random forest regressors (\Cref{app_sec:det_rf_regression}), and an adapted version of our probabilistic regularization scheme (\Cref{app_sec:det_ours_regression}).

\clearpage

\subsection{Regularized Linear Regression}
\label{app_sec:det_lin_reg_regression}

The \textbf{regularized linear regression models} were trained using the scikit-learn library \cite{scikit-learn}. The \textsc{liblinear} solver \cite{fan2008liblinear} was employed with a maximum of 1,000 iterations and a stopping tolerance of \num{1e-4}.
The models were independently fitted for all specified hyperparameter combinations presented in \Cref{tab:hyper_linreg_regression}.
The combination yielding the lowest validation set mean squared error was selected to evaluate the model on the held-out test set. 

\setlength{\tabcolsep}{18.0pt}
\begin{table}[H]
\centering
\caption{Hyperparameters for $L-1$ and $L_2$-Regularized Linear Regression}
\label{tab:hyper_linreg_regression}
\vspace{0.3em}
\begin{tabular}{lll}
\toprule
\textbf{Model} & \textbf{Hyperparameter} & \textbf{Search Space} \\
\midrule
Linear Regression & regularization type & $\ell1, \ell2$ \\
& regularization strength & 100 values spaced log-linearly in [\num{1e-4}, \num{1e4}] \\
\bottomrule
\end{tabular}
\end{table}

\subsection{Random Forest Regressors}
\label{app_sec:det_rf_regression}

The \textbf{random forest regression models} were trained using the scikit-learn library \cite{scikit-learn}. The models consisted of 100 decision trees with the \textsc{gini} splitting criterion. Each model was independently fitted for all specified hyperparameter combinations shown in~\Cref{tab:rfr_hypers_regression}. The combination with the lowest validation set mean squared error was selected to evaluate the model on the held-out test set.

\setlength{\tabcolsep}{32.0pt}
\begin{table}[H]
\centering
\caption{Hyperparameters for Random Forest Regressor}
\label{tab:rfr_hypers_regression}
\vspace{0.3em}
\begin{tabular}{lll}
\toprule
\textbf{Model} & \textbf{Hyperparameter} & \textbf{Search Space} \\
\midrule
Random Forest & maximum depth & 50 values spaced linearly in [5, 500] \\
& min. samples per split & 5, 15, 50, 100 \\
& min. samples per leaf & 1, 5, 10, 30, 100 \\
\bottomrule
\end{tabular}
\end{table}

\subsection{Q-SAVI}
\label{app_sec:det_ours_regression}

The regression variant of our probabilistic regularization scheme was set up identically to the classification variant described in~\Cref{app_sec:det_ours}, the only difference being the likelihood function used to evaluate~\Cref{eq:objective}. Instead of specifying $\log p(\by_{\calD} \vbar f(\bx_{\calD} ; \btheta))$ as a Bernoulli likelihood, a homoscedastic multivariate Normal likelihood with a unit diagonal covariance matrix was used. While a more expressive approach of either optimizing the covariance as a hyperparameter or letting the network predict point-wise means and variances to use in combination with a heteroscedastic likelihood function is possible, this straightforward method was found to be sufficient in this context. The full hyperparameter search space is presented in~\Cref{tab:ours_hypers_regression} and is identical to that in ~\Cref{tab:ours_hypers}.

\begin{table}[H]
\centering
\caption{Hyperparameters for Our Model}
\label{tab:ours_hypers_regression}
\vspace{0.3em}
\begin{tabular}{lll}
\toprule
\textbf{Model} & \textbf{Hyperparameter} & \textbf{Search Space} \\
\midrule
Ours & learning rate & \num{1e-4}, \num{1e-3} \\
 & number of layers & 2, 4, 6 \\
 & embedding dimension & 32, 64 \\
 & prior variance & \num{1e-2}, \num{1e-1}, \num{1e0}, \num{1e1}, \num{1e2} \\
 & context points per sample & 16, 128 \\
\bottomrule
\end{tabular}
\end{table}

\end{appendices}

\end{document}